\documentclass[11pt]{article}
\pdfoutput=1
\usepackage{jcapmod}

\usepackage{shorthand}
\usepackage{mathtools}
\usepackage{booktabs}
\usepackage[english]{babel}
\usepackage{amsmath,amssymb,amsbsy,amstext, amsthm, simplewick, amsfonts}
\usepackage{hyperref}
\usepackage{graphicx}
\usepackage[small]{caption}
\usepackage{siunitx}
\usepackage{upgreek}
\usepackage{framed}
\usepackage{wrapfig}
\usepackage{multirow}
\usepackage{bbm}
\usepackage[svgnames,dvipsnames,x11names]{xcolor}

\usepackage{selinput}

\usepackage{bm}
\usepackage{float}
\usepackage{geometry}
\usepackage{yfonts}
\usepackage{caption}
\usepackage{subcaption}
\usepackage{sidecap}
\usepackage{longtable}
\usepackage{anyfontsize}
\usepackage{dsfont}
\usepackage{tikz}
\usepackage{relsize}
\usepackage{tcolorbox}

\usepackage{xcolor}
\usepackage{xparse}

\usepackage{slashed}
\usepackage{simpler-wick}

\NewDocumentCommand{\colornucleus}{omme{_^}}{%
  \begingroup\colorlet{currcolor}{.}%
  \IfValueTF{#1}
   {\textcolor[#1]{#2}}
   {\textcolor{#2}}
    {%
     #3
     \IfValueT{#4}{_{\textcolor{currcolor}{#4}}}
     \IfValueT{#5}{^{\textcolor{currcolor}{#5}}}
    }%
  \endgroup
}

\usepackage{array}
\newcolumntype{L}[1]{>{\raggedright\let\newline\\\arraybackslash\hspace{0pt}}m{#1}}
\newcolumntype{C}[1]{>{\centering\let\newline\\\arraybackslash\hspace{0pt}}m{#1}}
\newcolumntype{R}[1]{>{\raggedleft\let\newline\\\arraybackslash\hspace{0pt}}m{#1}}

\usepackage[framemethod=default]{mdframed}
\newmdenv[skipabove=7pt,
skipbelow=7pt,
rightline=false,
leftline=false,
topline=false,
bottomline=false,
backgroundcolor=gray!10,
linecolor=gray,
innerleftmargin=5pt,
innerrightmargin=5pt,
innertopmargin=5pt,
innerbottommargin=5pt,
leftmargin=0cm,
rightmargin=0cm,
linewidth=4pt]{eBox}
\newmdenv[skipabove=7pt,
skipbelow=7pt,
rightline=false,
leftline=false,
topline=false,
bottomline=false,
backgroundcolor=gray!10,
linecolor=gray,
innerleftmargin=5pt,
innerrightmargin=5pt,
innertopmargin=-5pt,
innerbottommargin=5pt,
leftmargin=0cm,
rightmargin=0cm,
linewidth=4pt]{eBox2}
\usepackage[framemethod=default]{mdframed}
\newmdenv[skipabove=7pt,
skipbelow=7pt,
rightline=true,
leftline=true,
topline=true,
bottomline=true,
backgroundcolor=gray!15,
linecolor=gray,
innerleftmargin=5pt,
innerrightmargin=5pt,
innertopmargin=5pt,
innerbottommargin=5pt,
leftmargin=0cm,
rightmargin=0cm,
linewidth=0.75pt]{eBox3}

\definecolor{Red}{RGB}{214, 39, 40}
\definecolor{Blue}{RGB} {31, 119, 180}
\definecolor{Orange}{RGB}{255, 153, 51}
\definecolor{Purple}{RGB}{178, 102, 255}
\definecolor{Green}{RGB}{44, 160, 44}

\definecolor{vio}{RGB}{19, 130, 164}
\definecolor{vioo}{RGB}{89, 2, 155}
\newcommand{\Comment}[1]{{}}
\definecolor{darkblue}{rgb}{0.15,0.35,0.55}
\definecolor{reddish}{rgb}{0.65, 0.2, 0.2}
\definecolor{darkgreen}{RGB}{50,150,0}
\definecolor{greyish}{rgb}{.90,.90,.90}
\definecolor{greyish2}{rgb}{.96,.96,.96}
\definecolor{greyish3}{rgb}{.37,.37,.37}
\definecolor{darkblue2}{rgb}{0.3,0.4,0.9}
\definecolor{Blue3}{RGB}{31, 119, 180}
\usepackage[linktocpage=true]{hyperref}
\hypersetup{
colorlinks=true,
citecolor=darkblue,
linkcolor=reddish,
urlcolor=darkblue,
pdfauthor={},
pdftitle={},
pdfsubject={}
}

\usepackage{colortbl}
\definecolor{lightgreen}{cmyk}{0.2, 0, 0.2, 0.2}
\definecolor{lightgray2}{cmyk}{0.1,0.1,0,0.1}
\definecolor{Red2}{RGB}{214, 39, 40}
\definecolor{Blue2}{RGB} {31, 119, 180}
\definecolor{Orange2}{RGB}{255, 127, 14}
\definecolor{Green2}{RGB}{44, 160, 44}

\makeatletter
\newlength{\apb@width}
\newcommand{\autoparbox}[2][c]{\settowidth{\apb@width}{#2}\parbox[#1]{\apb@width}{#2}}

\makeatother


\def\hs{\hskip 1pt}

\def\beq{\begin{equation}}
\def\eeq{\end{equation}}
\def\be{\begin{equation}}
\def\ee{\end{equation}}

\newcommand{\ud}{{\rm d}}

\newcommand{\dif}{\mathrm{d}}

\allowdisplaybreaks[1]
\begin{document}

\newgeometry{top=2cm, bottom=2cm, left=2cm, right=2cm}

\begin{titlepage}
\setcounter{page}{1} \baselineskip=15.5pt 
\thispagestyle{empty}

\begin{center}
{\fontsize{21}{18} \bf The Cosmological Grassmannian }
\end{center}

\vskip 20pt
\begin{center}
\noindent
{\fontsize{14}{18}\selectfont 
Mattia Arundine\hs$^{1,2}$, Daniel Baumann\hs$^{1,2,3}$,  Mang Hei Gordon Lee\hs$^{2}$,\\[8pt]
Guilherme L.~Pimentel\hs$^{4}$  and Facundo Rost\hs$^{4}$}
\end{center}

\begin{center}
 \vskip8pt
\textit{$^1$ Institute of Physics, University of Amsterdam, Amsterdam, 1098 XH, The Netherlands}

  \vskip8pt
\textit{$^2$  Leung Center for Cosmology and Particle Astrophysics,
Taipei 10617, Taiwan}

  \vskip8pt
\textit{$^3$  Max Planck--IAS--NTU Center for Particle Physics, Cosmology and Geometry,
Taipei 10617, Taiwan}
\vskip 8pt
\textit{$^4$ Scuola Normale Superiore and INFN, Piazza dei Cavalieri 7, 56126, Pisa, Italy}
\end{center}

\vspace{0.4cm}
\begin{center}{\bf Abstract}
\end{center}
\noindent
We introduce the orthogonal Grassmannian as a novel kinematic space for describing correlators of massless spinning fields in de Sitter space. By automatically encoding the constraints of conformal symmetry and current conservation, the formalism drastically simplifies these correlators.   We show that three-point functions are fixed by little group covariance and take the same form as the corresponding Schwinger-parameterized correlators in twistor space. 
The power of the Grassmannian approach is especially evident for four-point functions, which require dynamical input beyond kinematics. We demonstrate that unitarity enforces the same factorization properties as for scattering amplitudes and use these to bootstrap the four-point functions in several non-trivial examples, including Yang--Mills theory and gravity. We find expressions that are astonishingly simple and reveal a close connection to the corresponding scattering amplitudes. 
Our results suggest that the Grassmannian provides the natural language for spinning correlators in de Sitter space and illuminates their geometric origin.

\end{titlepage}
\restoregeometry

\newpage
\setcounter{tocdepth}{3}
\setcounter{page}{2}

\linespread{0.95}
\tableofcontents
\linespread{1.1}

\newpage

\section{Introduction}

Unitarity, locality and Lorentz symmetry impose stringent constraints on the dynamics of elementary particles in flat space. These constraints are especially restrictive for massless particles with spin, which dictate the laws of physics at long distances. The allowed interactions of massless spin-1 particles must take the form of Yang–Mills theory, while those of massless spin-2 particles lead to Einstein gravity with a universal coupling to matter~\cite{Benincasa:2007xk,Wald:1986bj}. Forces mediated by massless higher-spin particles do not exist~\cite{Weinberg:1965nx,Benincasa:2007xk,Wald:1986bj, Weinberg:1980kq,Porrati:2008rm,McGady:2013sga}. Even when the dynamical theory exists, making locality manifest requires introducing unphysical gauge degrees of freedom, which cancel in physical observables. This leads to an unreasonable complexity in the perturbative computations of massless scattering amplitudes. 

\vskip 4pt
One of the lessons of the modern S-matrix bootstrap is that it pays to focus directly on physical (``on-shell") observables and to adopt a language that, while making the requirements of quantum mechanics and relativity manifest, carries no extra baggage. The reason is twofold: first, the resulting scattering amplitudes are often far simpler than the multitude of Feynman diagrams from which they are assembled~\cite{Parke:1986gb}; second, the striking simplicity of the final results hints at underlying mathematical structures that effectively ``compute" them directly~\cite{Arkani-Hamed:2013jha}. At a practical level, this on-shell philosophy also leads to substantial simplifications in the computation of scattering amplitudes, yielding ``theoretical data" that can be mined for additional mathematical structures and new physical insights.

\vskip 4pt
In cosmology, the situation is more prehistorical, resembling our limited understanding of scattering amplitudes before the on-shell revolution. 
This is especially true for massless theories with spin.
Already at four points, explicit expressions for correlators in gauge theories and gravity are frighteningly complex in momentum space~\cite{Baumann:2020dch, Baumann:2021fxj, Albayrak:2020fyp, Bonifacio:2022vwa}.
Moreover, while the generalization of spinor helicity variables from flat space to de Sitter space is useful at three points, it becomes very cumbersome at four points. 
Recently, it was argued that twistors are a more natural language for massless correlators  because, unlike cosmological spinor helicity variables, they make all kinematic constraints manifest~\cite{Baumann:2024ttn}  (see also~\cite{Bala:2025gmz, Bala:2025qxr, S:2025pmh, CarrilloGonzalez:2025qjk, Ansari:2025fvi, Bala:2025jbh}). In particular, twistors have a built-in ``holomorphicity” which  imposes that massless particles propagate the right number of degrees of freedom required by unitarity. In that sense, twistors are the (Anti-)de Sitter analogue of spinor helicity variables for massless particles in flat space. 
 In~\cite{Baumann:2024ttn}, the power of twistors  was demonstrated explicitly for three-point functions, which are completely fixed by kinematics and  take a remarkably simple form in twistor space. 
An important challenge is to extend this treatment to higher points, where additional dynamical input (beyond pure kinematics) is needed.

\vskip 4pt
An important property of correlators (and amplitudes) in  twistor space is that they are distributional, i.e.~they are composed of delta functions and their derivatives. This makes it more complicated to manipulate them analytically and to bootstrap anything beyond three-point functions.  A promising way to proceed is to ``Schwinger parameterize” the distributions~\cite{Arkani-Hamed:2009hub,Arkani-Hamed:2009ljj}.  The Schwinger parameters $c_{ij}$ are conjugate variables that arise from the Fourier transform of the inner products  $Z_i \cdot Z_j$, where $Z_i$ are the twistor coordinates associated with the fields $i=1,\cdots\hspace{-1pt}, n$.  One finds that the resulting object is a rational function exhibiting factorization properties closely analogous to those of scattering amplitudes. 

\vskip 4pt
As we will show in this paper, the Schwinger parameters $c_{ij}$ acquire a life of their own as coordinates on the {\it orthogonal Grassmannian}  ${\rm OGr}(n,2n)$, which we refer to as the ``cosmological Grassmannian.''
 We will demonstrate how four-point functions can be bootstrapped directly in Grassmannian space by exploiting the factorization properties of the correlators. Remarkably, the resulting Grassmannian correlators are extraordinarily simple (nearly as simple as their flat-space scattering-amplitude counterparts) thereby revealing, for the first time, the hidden simplicity of spinning cosmological correlators.

\vskip 4pt
The Grassmannian structure governing spinning correlators in de Sitter space can also be discovered without introducing twistors. 
This begins with the observation that cosmological spinor helicity variables fail at the crucial task of enforcing de Sitter boost isometries (or, equivalently, special conformal transformations on the boundary). As we will see below, these constraints imply that the correlators must solve a Laplace-like equation in the spinor variables. Of course, the Laplace equation in two dimensions is solved trivially by (anti-)holomorphic functions of the complex coordinates $x \pm i y$. This starts to rhyme with the song that led us to twistors. Amazingly, the Grassmannian generalizes a recipe to build holomorphic combinations of variables that solve the (even-dimensional) Laplace equation. 
The Grassmannian therefore resolves the final remaining hurdle encountered in the spinor helicity formalism.

\vskip 4pt
We find it intriguing that the Grassmannian makes its first appearance in a cosmological setting, given its central role in the modern geometric formulation of scattering amplitudes. 
Famously, in planar $\mathcal{N}=4$ supersymmetric Yang--Mills theory, the scattering amplitudes of $n$ particles, with $k$ negative helicities, are encoded as integrals over the Grassmannian ${\rm Gr}(k,n)$, the space of  $k$-dimensional planes in $n$-dimensional space~\cite{Arkani-Hamed:2009ljj, Arkani-Hamed:2009pfk}. Each physical amplitude corresponds to a specific contour integral of a canonical differential form on this space. The poles of the integrand encode kinematic constraints such as momentum conservation and on-shell conditions, while residues at these poles correspond to individual on-shell diagrams~\cite{Arkani-Hamed:2012zlh}.  Beyond providing a compact representation of amplitudes, the Grassmannian framework also revealed deep geometric and combinatorial structures underlying scattering processes, ultimately culminating in the discovery of the {\it amplituhedron}~\cite{Arkani-Hamed:2013jha}. 

\vskip 4pt
Seen in this light, the emergence of a cosmological Grassmannian suggests that spinning correlators in de Sitter space may admit an equally geometric and on-shell formulation, pointing toward a unified Grassmannian language for flat-space amplitudes and cosmological observables. An important difference between the cosmological Grassmannian and its flat-space counterpart is that the latter makes all bulk conformal symmetries manifest and is therefore only naturally suited to conformal theories such as ${\cal N}=4$ super-Yang--Mills theory. The cosmological Grassmannian, by contrast, trivializes only the (A)dS isometries and thus provides a natural language for any massless quantum field theories in curved spacetime.
Notably, it can describe de Sitter correlators even for non-conformal theories, greatly expanding its domain of applicability.

\newpage
\subsubsection*{Strategy and Summary} 

The route to the ``cosmological Grassmannian" is conceptually very simple. In the following, we sketch the basic elements of the Grassmannian formalism, while reserving technical details and concrete examples for the main text.

\vskip 4pt
We study correlators of massless spin-$\ell$ fields on the future boundary of de Sitter spacetime. The information contained in these correlators is captured by $n$-point wavefunction coefficients~$\psi_n$, which can be expressed as correlators of dual conserved currents $J^{\mu_1 \cdots \mu_\ell}$. The kinematics of these wavefunction coefficients is conveniently described by cosmological spinor helicity variables~$\{\lambda_i^\alpha,\bar\lambda_i^\alpha\}$, with $\alpha=1,2$, for each particle $i=1,\cdots\hskip -1pt, n$~\cite{Maldacena:2011nz}. 
We package these $2n$  spinors in a $2n\times 2$ matrix $\Lambda^{I \alpha}\equiv\left(\lambda_i^\alpha,\bar\lambda_i^\alpha\right)^T$, where $I=1,\cdots\hskip -1pt, 2n$.

\vskip 4pt
Momentum conservation implies a quadratic algebraic constraint on the spinor helicity variables, $\Lambda^T \cdot Q\cdot \Lambda= 0$, while conformal invariance (and current conservation)  
leads to a quadratic differential  constraint on the wavefunction coefficients:\footnote{The constraint in (\ref{equ:Conformal}) is a combination of the conformal Ward identity and the Ward--Takahashi identity~\cite{Witten:2003nn, Bargheer:2009qu, Maldacena:2011nz}. The latter implies that the right-hand side should in fact be non-zero, proportional to the lower-point function~$\psi_{n-1}$. By solving the simpler {\it homogeneous} constraint (\ref{equ:Conformal}), we only derive a certain {\it discontinuity} of the correlator. The full correlator can, however, be recovered by a dispersive integral (see Section~\ref{ssec:dispersive}). \label{foot:1}}
\be
\left(\frac{\partial}{\partial \Lambda}\right)^T\cdot Q\cdot \frac{\partial}{\partial \Lambda}  \ \psi_n(\Lambda) = 0\,,  \qquad Q\equiv \begin{pmatrix}
				0& 1_{n\times n}\\
				1_{n\times n}&0
			\end{pmatrix} .
			\label{equ:Conformal}
\ee
The latter constraint is automatically satisfied by any function that depends on $\Lambda$ only through the combination $(C\cdot \Lambda)_a{}^\alpha = C_{aI} \Lambda^{I\alpha}$, where $C_{aI}$ is an $n\times 2n$ matrix obeying $C\cdot Q\cdot C^T=0$.   This defines a notion of {\it holomorphicity} for the solutions. Momentum conservation, furthermore,
requires that this holomorphic function only has support at $C\cdot \Lambda = 0$.

\vskip 4pt
The auxiliary matrix $C$ is defined up to left multiplication by $\mathrm{GL}(n)$, $C \mapsto R\cdot C$, and the resulting equivalence classes define the {\it orthogonal Grassmannian} $\mathrm{OGr}(n,2n)$.\footnote{The orthogonal Grassmannian first appeared in the study of three-dimensional scattering amplitudes in $\mathcal{N}=6$ ABJM theory, where the kinematics naturally respect a three-dimensional conformal symmetry~\cite{Huang:2013owa,Kim:2014hva}. It has since also found applications in descriptions of the Ising model~\cite{Galashin:2020jvd}.} 
We can fix the GL$(n)$ redundancy by picking a gauge for the matrix $C$. For example, we may choose  $C =\big(1_{n\times n}\hs ,\hs C_n\big)$, 
where $C_n$ is a $n\times n$ matrix with elements $(C_n)_{ij} \equiv -c_{ij}$.  The parameters $c_{ij}$ then play the same role as the Schwinger parameters introduced in~\cite{Baumann:2024ttn} when formulating the correlators of conserved currents in twistor space (see Appendix~\ref{app:twistors}).

\begin{figure}[t!]
	\centering
	\includegraphics[scale=0.5]{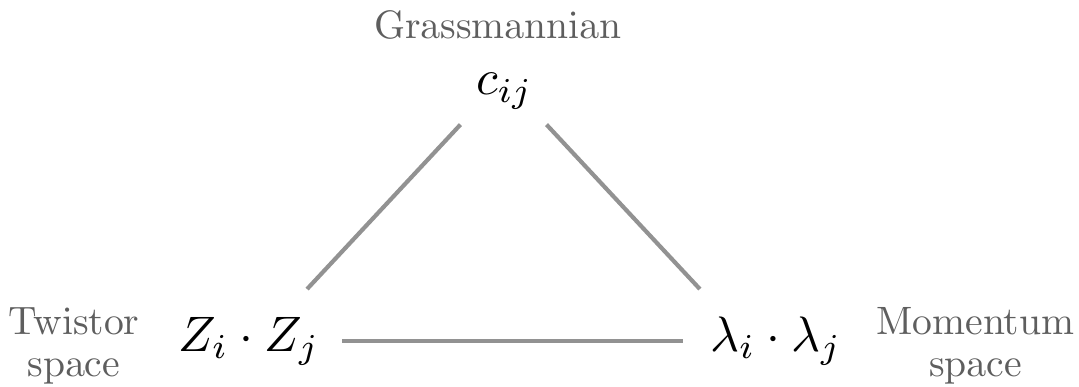} 
	\caption{Illustration of the different spaces used in this paper. Twistor space is related to momentum space by a half-Fourier transform and to the Grassmannian space by an ordinary Fourier transform. The integral transform in \eqref{equ:main-integral} maps Grassmannian correlators directly to momentum space.}
	\label{fig:spaces}
\end{figure}

\vskip 4pt
To remove the dependence on the arbitrary choice of $C$, we integrate over the coordinates of the Grassmannian, obtaining the following expression for the wavefunction coefficients: 
\beq
\psi_n(\Lambda) =\int {\rm d} C\, \delta(C\cdot \Lambda)\, A_n(C) \,,
\label{equ:main-integral}
\eeq
where consistency with the $\mathrm{GL}(n)$ redundancy requires that the function $A_n$ depends only on the {\it minors} of $C$, i.e.~$(I_1\cdots I_n) \equiv \epsilon^{a_1 \cdots a_n} C_{a_1 I_1} \cdots C_{a_n I_n}$.   We will show that the functions $A_n(C)$ take a particularly simple form, revealing a hidden simplicity of the correlators when expressed in Grassmannian space.

\vskip 4pt
The three-point functions $A_3$ can be bootstrapped using little group covariance alone. 
For example, the $+++$ correlator of a spin-$\ell$ field is
\beq
A_{3,+++} = \left(\frac{(\bar 1 \bar 2 \bar 3)^2}{{\cal K}} \right)^\ell \,,
\label{equ:A3-result}
\eeq
where ${\cal K} \equiv( 1 \bar1 2)(\bar 2 3\bar 3)$ is  a little group-invariant factor. Correlators with different helicities
 are obtained simply by interchanging barred and unbarred columns in the minor $(\bar 1 \bar 2 \bar 3)$. After gauge fixing, we get $A_{3,+++}=  \left( c_{12}\hs c_{23}\hs c_{31} \right)^{-\ell}$, which is of the same form as the Schwinger-parameterized correlator derived in twistor space~\cite{Baumann:2024ttn}.
In the limit ${\cal K} \to 0$, the result (\ref{equ:A3-result}) reproduces the correct flat-space amplitude.

\vskip 4pt
To describe the four-point functions $A_4$, we introduce the following minors
\beq
	S\equiv  (\bar 1\bar 212 )\,,\quad
		T \equiv (\bar 1\bar 414 )\,,\quad
U \equiv (\bar 1\bar 313 )\,,
\label{equ:Mandelstams}
\eeq
which are similar to the Mandelstam variables for amplitudes. 
Unlike in scattering amplitudes, however, the sum of these ``Mandelstams" doesn't vanish, i.e.~$S+T+U \ne 0$. 
Nevertheless, as in amplitudes, the exchange of massless particles produces poles at $S=0$, $T=0$, $U=0$, 
and unitarity requires the function $A_4$ to factorize on these poles (see Appendix~\ref{app:factorization}). 
Remarkably, within the Grassmannian formalism this factorization property is virtually the same as for amplitudes.
For example, the residue of the $s$-channel singularity is
\be \label{equ:Res-S=0-intro}
\mathop{\mathrm{Res}}_{S=0} A_4= \sum_hA_{3,L}^{(-h)} \hs A_{3,R}^{(+h)}\,,
\ee
where $A_{3,L}^{(-h)}$ and $A_{3,R}^{(+h)}$ are the relevant three-point Grassmannian correlators and $h$ is the helicity of the exchanged particle.
Imposing consistent factorization in all channels then puts important constraints on the allowed functions $A_4$. We will present many non-trivial examples in the paper.  For example, the (reduced) color-ordered all-plus correlator in pure Yang--Mills theory is\hs\footnote{As we explain in Section~\ref{sec:YM}, this result is only a part of the complete color-ordered correlator. In particular, it does not satisfy the Kleiss--Kuijf relation~\cite{Kleiss:1988ne}. The complete result is given in~\eqref{equ:AYM-kk}.} 
\beq
\hat A_{4,++++} =  2 \,\frac{(\bar 1 \bar 2 \bar 3 \bar 4)^2}{S T} \left(\frac{1}{S+T+U} + \frac{1}{S+T-U}\right)   .
\label{equ:A4-result}
\eeq
This is a remarkably simple expression, especially when compared to the complexity of its momentum-space counterpart~\cite{Albayrak:2018tam, Baumann:2020dch}. Note that the first term in (\ref{equ:A4-result}) is $(S+T+U)^{-1}$ times a function that looks like the flat-space amplitude when written in terms of the Mandelstams~(\ref{equ:Mandelstams}), while the second term arises from the interchange $U \to -\hs U$.  This intriguing structure of the Grassmannian correlator will be explored further in the main text.

\vskip 4pt
Recall that the integral~\eqref{equ:main-integral} solves the {\it homogeneous} constraint~\eqref{equ:Conformal} (see Footnote~\ref{foot:1}) and therefore only produces a certain {\it discontinuity} of the correlator, with the precise result depending on the choice of integration contour.  However, by choosing an integration contour that has the correct flat-space and factorization limits, we are able to find a specific discontinuity ${\rm Disc}[\psi_4]$ (with respect to the external energies), from which the full correlator $\psi_4$---that is, a solution of the {\it inhomogeneous} Ward identity---can be reconstructed via a dispersive integral (see Section~\ref{ssec:dispersive}). 
Note that our ${\rm Disc}[\psi_4]$ is distinct from the discontinuity that was recently studied in~\cite{Ansari:2025fvi}.
In the latter case, the discontinuity is taken with respect to the {\it internal} energy and therefore yields Wightman functions, which miss contributions from contact terms.

\subsubsection*{Outline} The paper is organized as follows. In Section~\ref{sec:Kinematics}, we review the kinematics of boundary correlators in de Sitter space and introduce the orthogonal Grassmannian as a powerful way to trivialize all kinematic constraints for massless spinning correlators. We show that the wavefunction coefficients can be written as an integral over the coordinates of the Grassmannian; cf.~(\ref{equ:main-integral}).
In Section~\ref{sec:3pt}, we use little group covariance to bootstrap three-point functions for fields with arbitrary spins, making contact with the twistor analysis in~\cite{Baumann:2024ttn}.
In Section~\ref{sec:4pt}, we introduce unitarity and the corresponding factorization limits as a key tool for bootstrapping four-point functions. 
We illustrate the Grassmannian formalism for the case of scalars exchanging a photon and a graviton and show that the known momentum-space answers are obtained by explicitly performing the integral in~(\ref{equ:main-integral}). In Section~\ref{sec:YM}, we present in detail the case of pure Yang--Mills theory. 
We show that the Grassmannian integral \eqref{equ:main-integral} computes a specific discontinuity of the momentum-space correlator, but that the full correlator can be reconstructed by performing an additional dispersive integral. Our conclusions are presented in Section~\ref{sec:conclusions}. 

\vskip 4pt
Four appendices contain supplemental material: In Appendix~\ref{app:twistors}, we give a brief introduction to twistors and review their recent application to spinning correlators in de Sitter space~\cite{Baumann:2024ttn}.
In Appendix~\ref{app:factorization}, we derive the factorization property used in Sections~\ref{sec:4pt} and~\ref{sec:YM}. 
In Appendix~\ref{app:flat}, we analyze the flat-space limit, demonstrating explicitly how the flat-space Grassmannian emerges from the cosmological Grassmannian and how, in this limit, the momentum-space correlators develop the expected total energy singularities.
Finally, in Appendix~\ref{app:gravity}, we derive the graviton four-point function in Grassmannian space. We show that this leads to a triple discontinuity of the wavefunction coefficient in momentum space, but don't perform the additional dispersive integral to reproduce the full correlator.

\subsubsection*{Notation}

Throughout the paper, we use natural units, $\hbar = c \equiv 1$, and the mostly plus convention for the metric. 
Although our narrative focuses mostly on de Sitter (dS) space, everything we say also holds for spinning correlators in Anti-de Sitter (AdS) space and the associated Lorentzian boundary conformal field theories (CFTs). 
We work in four bulk dimensions with a three-dimensional boundary, which is Euclidean in dS and Lorentzian in AdS.
To unify our treatment, we denote the boundary coordinates by $x^\mu$,  with $\mu=1,2,3$ (dS) or $\mu =0,1,2$ (AdS). The bulk coordinates are $X^M$, with $M=0,1,2,3$.

\vskip 4pt
We denote a 
spin-$\ell$ bulk field by $\Phi_{M_1 \cdots M_\ell}$ and its
boundary value by $\phi_{\mu_1 \cdots \mu_\ell}$.
The dual boundary current is  $J^{\mu_1 \cdots \mu_\ell}$, with $\partial_\mu J^{\mu \mu_2\cdots \mu_\ell}=0$.
We will often use the index-free notation $J_i^\pm \equiv \epsilon_{i,\mu_1}^\pm \cdots \epsilon_{i,\mu_{\ell_i}}^\pm J_i^{\mu_1\cdots \mu_{\ell_i}}(k_i^\mu)$, where $\epsilon_{i,\mu}^\pm$ are polarization vectors. This both eliminates clutter in our expressions and allows us to define fixed-helicity correlators by picking appropriate polarization vectors. 
We study $n$-point wavefunction coefficients on the boundary, which can be expressed as correlation functions of the dual currents:
\beq
	\psi_n(\underline{k}^\mu) \equiv \langle J_1^\pm \cdots J_n^\pm \rangle\,, 
\eeq
where $\underline{k}^\mu \equiv \{ k_1^\mu, \cdots\hspace{-1pt}, k_n^\mu\}$ is the set of boundary momenta. 
The momenta $k_i^\mu$ can be expressed in terms of the spinor helicity variables $\lambda_i^\alpha$ and $\bar \lambda_i^\beta$.  We raise and lower the spinor indices using the left multiplication convention, $\lambda_\alpha=\epsilon_{\alpha\beta}\lambda^\beta$ and $\lambda^\alpha=\epsilon^{\alpha\beta}\lambda_\beta$, with $\epsilon_{21}=\epsilon^{12} \equiv 1$.

\section{Kinematics and Grassmannian}
\label{sec:Kinematics}

We begin by showing how to trivialize the kinematic constraints of $n$-point correlators involving massless spinning fields on the future boundary of de Sitter space. 
We proceed in two steps. First, we introduce spinor helicity variables to express these correlators in a helicity basis. Next, we construct linear combinations of these variables that simultaneously linearize the constraints imposed by special conformal transformations and momentum conservation. We find that these linear combinations contain auxiliary variables 
 that live in the orthogonal Grassmannian.  The momentum-space correlators are then written as integrals over this Grassmannian space.

\subsection{Conserved Currents}
\label{sec:current}

Using the flat slicing, the line element of de Sitter spacetime is
\beq
{\rm d} s^2 = \frac{L^2}{\eta^2} \big(- {\rm d} \eta^2 + {\rm d} x^2\big) \, ,
\eeq
where $-\infty < \eta < 0$ is conformal time and $L$ is the curvature scale. We are interested in correlators defined on the $d$-dimensional boundary at $\eta \approx 0$. The boundary coordinates are~$x^\mu$, with $\mu=1,\cdots\hskip -1pt, d$. Bosonic particles with spin $\ell$ are described by symmetric tensor fields $\Phi_{M_1 \cdots M_\ell}$, which are
both transverse ($\nabla^M \Phi_{M M_2 \cdots M_\ell}=0$) and traceless ($\Phi^M{}_{MM_2 \cdots M_\ell}=0$).
In an appropriate gauge, the components of the field transverse to the boundary vanish ($\Phi_{0M_2 \cdots M_\ell}=0$) and  the components along the boundary $\Phi_{\mu_1 \cdots \mu_\ell}$ can be taken to be transverse and traceless. Near the $d$-dimensional boundary, this transverse-traceless spatial part of the field behaves as
\beq
\Phi_{\mu_1 \cdots \mu_\ell}(\eta,x^\mu) \xrightarrow{\ \eta \to 0\ } \phi_{\mu_1 \cdots \mu_\ell}(x^\mu)\hs \eta^{\bar \Delta - \ell}  + j_{\mu_1 \cdots \mu_\ell}(x^\mu)\hs \eta^{\Delta -\ell}\,,
\eeq
where $\bar \Delta$ is the scaling dimension of the field and $\Delta \equiv d- \bar \Delta$. For massless particles of spin $\ell$, we have $\bar \Delta = 2-\ell$ and $\Delta = \ell+d-2$. 
Note that, in the late-time limit $\eta \to 0$, the dominant mode is the gauge field $ \phi_{\mu_1 \cdots \mu_\ell}(x^\mu)$. Furthermore, the spacetime isometries act as conformal transformations on the boundary, under
which $\phi_{\mu_1 \cdots \mu_\ell}(x^\mu)$ transforms like a spin-$\ell$ conformal primary operator of weight~$\bar \Delta$.

\vskip 4pt
The statistics of the boundary field is described by a wavefunctional $\Psi[\phi]$, whose perturbative expansion is
\beq
\Psi[\phi] \approx \exp \left(- \sum_{n=2}^\infty \frac{1}{n!} \int {\rm d}^d x_1 \ldots {\rm d}^d x_n \, \psi_n(\underline{x}^\mu) \, \phi(x_1^\mu) \cdots \phi(x_n^\mu)\right) , 
\eeq
where we have suppressed the tensor indices.
The kernel functions $\psi_n(\underline{x}^\mu)$, with $\underline{x}^\mu \equiv \{ x_1^\mu, \cdots\hspace{-1pt}, x_n^\mu\}$, are the ``wavefunction coefficients"  and have the same kinematic properties as correlation functions in a conformal field theory
\beq
	\psi_n(\underline{x}^\mu) \equiv \langle J(x_1^\mu) \cdots J(x_n^\mu)\rangle\,, 
\label{equ:psi-n}
\eeq
with the operators $J(x^\mu)$ having conformal
dimensions $\Delta \equiv d - \bar \Delta$.  The wavefunction is invariant under gauge transformations of $\phi_{\mu_1\cdots \mu_\ell}(x^\mu)$ iff the dual operators $J^{\mu_1\cdots \mu_\ell}(x^\mu)$
are {\it conserved currents}, with  $\partial_\mu J^{\mu \mu_2 \cdots \mu_\ell} = 0$. The wavefunction coefficients  (\ref{equ:psi-n}) must then satisfy both the conformal Ward identities and the Ward--Takahashi identity of current conservation. As we will see, it is difficult to make both of these kinematic constraints manifest at the same time.

\subsection{Spinor Helicity Variables}

In momentum space, the wavefunction coefficients are functions of the boundary momenta $k^\mu_i$ and the external polarization vectors $\epsilon^\mu_i$. Alternatively, the kinematic data can also be written in terms of the cosmological spinor helicity variables introduced in~\cite{Maldacena:2011nz}. These can be defined by contracting the three-vector $k_i^\mu$ with the Pauli matrices $(\sigma^\mu)_\alpha^{~\beta}$ (for $\alpha,\beta=1,2$), and writing the result as an outer product of two-component spinors
\be \label{equ:shv}
k_{i,\mu} (\sigma^\mu)_{\alpha\beta}=\lambda_{i,(\alpha}\bar\lambda_{i,\beta)}\,,
\ee
where $(\sigma^\mu)_{\alpha\beta} \equiv \epsilon_{\beta\gamma}(\sigma^\mu)_\alpha^{~\gamma}$ is a symmetric $2\times 2$ matrix. 
Note that \eqref{equ:shv} is invariant under the little group transformation $\lambda_i\mapsto \rho_i\hs  \lambda_i$ and $\bar\lambda_i\mapsto \rho_i^{-1} \hs\bar\lambda_i$.\footnote{Once we impose a reality condition on the spinors, the little group transformation is restricted to either $\rho_i=e^{i\theta_i}\in \hs$U$(1)$ for Euclidean momenta, or $\rho_i\in \hs\mathbb{R}$ for spacelike Lorentzian momenta.}
We define spinor brackets as $\langle ij\rangle \equiv \epsilon_{\alpha\beta}\lambda_i^\alpha \hs \lambda_j^\beta$, $\langle \bar \imath \hs \bar \jmath\rangle \equiv \epsilon_{\alpha\beta}\bar\lambda_i^\alpha \hs \bar \lambda_j^\beta$ and $\langle i\hs \bar \jmath\rangle \equiv \epsilon_{\alpha\beta}\lambda_i^\alpha \hs \bar \lambda_j^\beta$.
Notice that, unlike in flat space, we are allowed to have mixed brackets between barred and unbarred spinors. In particular, the mixed bracket $\langle \bar \imath i\rangle=2k_i$ yields the energy $k_i\equiv |\vec k_i|$ of the leg $i$. 
		
	\vskip 4pt
	In a helicity basis, the wavefunction coefficients are conveniently written 
	in terms of these spinor helicity variables~$\{\lambda_{i}^\alpha,\bar\lambda_{i}^\beta\}$. For future convenience, we will normalize them as\footnote{Here, we use the index-free notation $J_i^\pm \equiv \epsilon_{i,\mu_1}^\pm \cdots \epsilon_{i,\mu_{\ell_i}}^\pm J_i^{\mu_1\cdots \mu_{\ell_i}}(k_i^\mu)$, where $\epsilon_{i,\mu}^\pm$ are polarization vectors.} 
	\be \label{equ:wvf-normalization}
	\psi_n(\lambda_{i}^\alpha,\bar\lambda_{i}^\beta) \equiv \left(\prod_{i=1}^n\frac{1}{k_i^{\Delta_i-2}}\right)\langle J_1^\pm \cdots J_n^\pm \rangle \,.
	\ee 
	These objects are covariant under the action of the {\it little group}: For particles with helicities $h_i=\pm \ell_i$, the little group scaling of $\psi_n$ is 
	\be \label{equ:littlegroup-psin}
	\psi_n(\rho_i \lambda_i^\alpha, \rho_i^{-1} \bar \lambda_i^\beta) = \left(\prod_{i=1}^n\rho_i^{-2 h_i}\right)\psi_n(\lambda_i^\alpha, \bar \lambda_i^\beta)\,.
	\ee 
Moreover, the boundary correlators in de Sitter space (and the associated wavefunction coefficients) are invariant under the conformal transformations of its three-dimensional boundary. 
\begin{itemize}
\item Invariance under {\it rotations}  implies that all spinor indices~$\alpha,\beta$ in $\psi_n$ must be contracted between each other, and thus $\psi_n$ will only depend on spinor brackets. 
\item Invariance under {\it translations} means that $\psi_n$ only has support when the sum of the spatial momenta vanishes
\be 
\sum_{i=1}^n k_i^\mu=0\iff \sum_{i=1}^n \left(\lambda_i^\alpha \bar\lambda_i^\beta +\lambda_i^\beta \bar\lambda_i^\alpha \right)=0\,.
\label{equ:momcons-shv}
\ee 
The function $\psi_n$  therefore contains a $\delta$-function of momentum conservation, $\psi_n \propto \delta(\sum_i \vec k_i)$.
\item Given the normalization in \eqref{equ:wvf-normalization},
 invariance under {\it dilatations} implies the scaling property
\be \label{equ:dil-2n}
\psi_n(r\lambda_i^\alpha, r\bar \lambda_i^\beta)=r^{-2n}\hs\psi_n(\lambda_{i}^\alpha,\bar\lambda_{i}^\beta) \,,
\ee 
where $r \in \mathbb{R}^+$. 
\item Finally, the most non-trivial constraint is invariance under {\it special conformal transformations}~\cite{Witten:2003nn,Baumann:2020dch,Baumann:2021fxj}.  In this paper, we will study solutions of the following differential constraint
\be 
\label{equ:sct-hom}
\sum_i \frac{\partial^2}{\partial \lambda_i^{(\alpha}\partial\bar \lambda_i^{\beta)}}\,\psi_n\,=\,
0\,.
\ee 
We will see that the solution of this constraint gives a certain {\it discontinuity} of the full correlator. The full correlator would solve a generalization of \eqref{equ:sct-hom} with a nonzero right-hand side proportional to the $(n-1)$-point correlator $\psi_{n-1}$~\cite{Baumann:2020dch,Baumann:2021fxj}.
We will recover the full correlator by integrating the discontinuity (see Section~\ref{ssec:dispersive}).
\end{itemize}
It is challenging to solve the two quadratic constraints~\eqref{equ:momcons-shv} and~\eqref{equ:sct-hom} simultaneously. In the following, we will show how the cosmological Grassmannian arises naturally when we attempt to make these symmetries manifest.

\subsection{Cosmological Grassmannian}

To discuss the solutions of the constraints~\eqref{equ:momcons-shv} and~\eqref{equ:sct-hom}, 
it is convenient  to package the $2n$ spinor helicity variables $\{\lambda_i^\alpha,\bar\lambda_i^\alpha\}$ in a $2n\times 2$ matrix 
\be \label{equ:BigLambda}
\Lambda \equiv\begin{pmatrix}
	\lambda_1^1 & \lambda_1^2\\
	\vdots & \vdots \\
	\lambda_n^1 & \lambda_n^2 \\[4pt]
	\bar\lambda_1^1 & \bar\lambda_1^2 \\
	\vdots &\vdots \\
	\bar\lambda_n^1 & \bar\lambda_n^2
\end{pmatrix} .
	\ee 
	We take $\Lambda$ to be real, so that the  three-momenta $k_i^\mu$ are spacelike on a $(2+1)$-dimensional Lorentzian boundary. Ultimately, our results are related by analytic continuation to those obtained for a theory on the Euclidean boundary of dS$_4$.

\vskip 4pt
The constraint of momentum conservation \eqref{equ:momcons-shv} then becomes 
	\be \label{equ:barLQL}
 \Lambda^T \cdot Q\cdot \Lambda= 0  \,, \qquad 	Q\equiv \begin{pmatrix}
				0& 1_{n\times n}\\
				1_{n\times n}&0
			\end{pmatrix} .
	\ee 
The kinematic data encoded in the matrix $\Lambda$ has a nice geometric interpretation. Note that each column in~\eqref{equ:BigLambda} defines a vector in a $2n$-dimensional space. Under a GL$(2)$ transformation of the form $\Lambda^{I\alpha}\mapsto \Lambda^{I\beta} R_\beta^{~\alpha}$, describing spatial rotations and dilatations, these two vectors change, but the $2$-plane spanned by them is invariant. Momentum conservation, in the form of the constraint~\eqref{equ:barLQL}, then implies that this $2$-plane is null (with respect to the metric $Q$).

\vskip 4pt	
Similarly, the SCT constraint \eqref{equ:sct-hom} can be expressed as 
\be \label{equ:sct-hom-indices}
\frac{\partial}{\partial\Lambda^{I\alpha}}\hs Q^{IJ}\hs\frac{\partial}{\partial\Lambda^{J\beta}}\, \psi(\Lambda)=0\,.
\ee 
We observe that both \eqref{equ:barLQL} and \eqref{equ:sct-hom-indices} are quadratic in the spinor helicity variables, making the constraints nonlinear and difficult to solve directly. To simplify the problem, we introduce auxiliary variables that render the constraints linear and simultaneously define a notion of holomorphicity for the solutions.
Concretely, we try an ansatz of the form
\be \label{equ:Fn-CL}
 \psi_n(\Lambda)=F_n(C\cdot \Lambda)\,,
\ee
where $C=C_{aI}$ is a matrix.  We now show explicitly how the constraints \eqref{equ:barLQL} and \eqref{equ:sct-hom-indices} impose specific conditions on the matrix $C$.

\vskip 4pt
First, we note that the SCT constraint~\eqref{equ:sct-hom-indices} now reads
\be 
\left(C_{aI}Q^{IJ}C_{bJ}\right) \frac{\partial^2}{\partial(C_{aK}\Lambda^{K\alpha})\partial(C_{bL}\Lambda^{L\beta})}F_n(C\cdot \Lambda)=0\,,
\ee 
which is trivially solved if $C$ satisfies
\be \label{equ:CQC2}
C\cdot Q\cdot C^T=0\,.
\ee  
Because of this constraint, the matrix $C_{aI}$ can have at most $n$ rows. Without loss of generality, we take it to be an $n\times 2n$ matrix, with rows labeled by $a=1,2,\cdots\hskip -1pt,n$. Each row of the matrix $C_{aI}$ defines a vector in a $2n$-dimensional space.
Under a GL$(n)$ transformation of the form $C \mapsto R \cdot C$, these vectors change, but the $n$-plane spanned by them is invariant. Geometrically, we can therefore associate an $n$-plane to each choice of the matrix $C$. Due to the constraint~\eqref{equ:CQC2}, each such $n$-plane is null, with respect to the metric $Q$. The space of null $n$-planes in $2n$ dimensions is the {\it orthogonal Grassmannian} OGr$(n,2n)$. More algebraically, a null $n$-plane picks out the subset of variables on which the function $F_n(C\cdot \Lambda)$ depends, which are half of the variables of~$\psi_n(\Lambda)$. Each null $n$-plane therefore defines a notion of holomorphicity that renders the Laplace-like equation~\eqref{equ:sct-hom-indices} automatic.\footnote{This generalizes the usual two-dimensional notion of {\it holomorphicity}: A null line in two dimensions, described by its direction $\vec c=(1,\pm i)$, picks out the only variable $\vec c\cdot \vec x$ on which the function~$f(\vec x)=F(\vec c\cdot \vec x)$ depends, with $\vec x=(x,y)$. This function solves the two-dimensional Laplace equation $\nabla^2 f=0$ automatically.}
 
 \vskip 4pt
The remaining non-trivial constraint is the momentum conservation in \eqref{equ:barLQL}, $\Lambda^T\cdot Q\cdot \Lambda=0$. Naively, it looks like this spoils holomorphicity because it seems to depend on $\Lambda$ not only through the combination $C\cdot \Lambda$. However, there is a simple way to linearize this quadratic constraint, while preserving holomorphicity, by imposing
\be \label{equ:CL}
C\cdot \Lambda=0\,.
\ee 
To see that this implies the momentum-conservation constraint~\eqref{equ:barLQL}, we first note that, because the matrix $C$ has has exactly $n$ rows,
 any $2n$-dimensional vector $\Lambda^{I\alpha}$ (with fixed $\alpha=1,2$) satisfying the $n$ equations $(C\cdot \Lambda)_a{}^\alpha=0$ can be expressed as a linear combination of the $n$ rows of the matrix $C\cdot Q$.  
Any solution of~\eqref{equ:CL} can therefore be written as
	$\Lambda=(C\cdot Q)^T \cdot P $, for some arbitrary $n\times 2$ matrix $P$. Substituting this into~\eqref{equ:barLQL} and using $Q^2 = Q$, we see that  momentum conservation follows from the orthogonality condition~\eqref{equ:CQC2}. 

\vskip 4pt
Hence, 
we have been led to the ansatz
\be\label{equ:deltaCL-A}
F_n(C\cdot \Lambda) = \delta(C\cdot \Lambda)\, A_n(C)\,,
\ee
where the $2n$-dimensional $\delta$-function enforces $C\cdot \Lambda=0$ and the function $A_n(C)$ allows for an additional explicit dependence on the matrix $C$. Note that spatial rotations act as SL$(2)$ transformations $\Lambda^{I\alpha}\mapsto \Lambda^{I\beta} R_\beta^{~\alpha}$, with $\det(R_\beta^{~\alpha})=1$, which leaves the ansatz  invariant. Moreover, 
the dilatation constraint \eqref{equ:dil-2n} is satisfied because \eqref{equ:deltaCL-A} scales as $r^{-2n}$ after a rescaling $\Lambda\mapsto r\hs \Lambda$.

\begin{figure}[t!]
	\centering
	\includegraphics[scale=1.2]{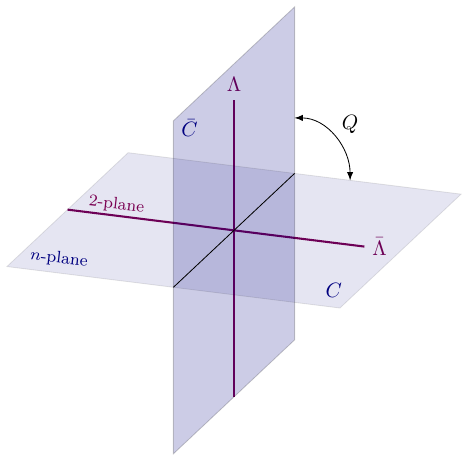} 
	\caption{Graphical illustration of the geometry underlying the cosmological Grassmannian. The Grassmannian matrix $C$ defines an $n$-plane in $2n$ dimensions. Conformal invariance requires this plane to be null with respect to the metric $Q$, or equivalently, orthogonal to the $n$-plane defined by the auxiliary matrix~$\bar C\equiv C\cdot Q$. The apparent intersection of the two planes is only an artifact of the lower-dimensional drawing. The spinor helicity variables $\Lambda$ define a $2$-plane. Momentum conservation implies that this $2$-plane is orthogonal to the $n$-plane corresponding to $C$ and therefore lies in $\bar C$.}
	\label{fig:planes}
\end{figure}

\vskip 4pt
The interplay between the Grassmannian and the spinor helicity variables admits a natural geometric interpretation, depicted in Figure~\ref{fig:planes}. First, recall that the orthogonal Grassmannian ${\rm OGr}(n,2n)$ is the space of null $n$-planes passing through the origin in a $2n$-dimensional space.
	Each matrix $C$ therefore defines such an $n$-plane.  
	Conformal invariance implies that the planes associated to $C$ and $\bar C \equiv C \cdot Q$ are orthogonal. Since the space is $2n$-dimensional, these two $n$-planes are orthogonal complements of each other. Similarly, the matrix $\Lambda$ defines a $2$-plane in the same $2n$-dimensional space. 
	The constraint $C\cdot \Lambda=0$ associated to momentum conservation means that this $2$-plane is orthogonal to the plane defined by $C$. Equivalently, this $2$-plane must lie within the $n$-plane associated to $\bar C$, as it is the orthogonal complement of the plane defined by $C$. 
	By the same reasoning, the auxiliary matrix $\bar\Lambda\equiv Q\cdot \Lambda$ defines a $2$-plane contained in the plane associated with $C$. As illustrated in Figure~\ref{fig:planes}, the two $2$-planes defined by $\Lambda$ and $\bar\Lambda$ are therefore contained in $\bar C$ and $C$, respectively, and are orthogonal to each other.

\subsection{Wavefunction Coefficients}
\label{ssec:WF}

We just showed how all isometries are trivialized by the holomorphic function~\eqref{equ:deltaCL-A}, at the expense of introducing the auxiliary matrix  $C$ in the orthogonal Grassmannian OGr$(n,2n)$. To erase the dependence on this arbitrary matrix, we integrate over it, yielding the following expression for the wavefunction coefficients in momentum space:
\beq\label{equ:grassmanian-integral}
 \psi_n(\Lambda) =\int {\rm d}C\, \delta(C\cdot \Lambda)\, A_n(C)\,.
\eeq
We will first explain the different elements of this formula and then apply it to some explicit examples.

\subsubsection*{GL$\boldsymbol{(n)}$ invariance} 

To begin with, let us clarify the integration measure $\ud C$ in \eqref{equ:grassmanian-integral}. It is the natural measure on the orthogonal Grassmannian $\mathrm{OGr}(n,2n)$, defined as
\be \label{equ:measure}
{\rm d} C=\frac{{\rm d}^{n\times 2n}C}{\text{GL}(n)} \, \delta(C\cdot Q\cdot C^T)\,,
\ee 
where we divide out the $\mathrm{GL}(n)$ redundancy $C\mapsto R\cdot C$ and impose the orthogonality constraint~\eqref{equ:CQC2} by means of a $\delta$-function. Consequently, the integral \eqref{equ:grassmanian-integral} is only well defined if the integrand is also invariant under $\mathrm{GL}(n)$.

\vskip 4pt
Invariance under the subgroup SL$(n)$ implies that the function $A_n(C)$ 
can depend on the matrix $C$ only through its {\it minors}
\be 
(I_1\cdots I_n)= \epsilon^{a_1\cdots a_n}C_{a_1I_1}\cdots C_{a_nI_n}\,,
\ee  
which are the determinants of the $n\times n$ sub-matrices of $C$ comprising the $n$ columns $I_1,\cdots\hskip -1pt,I_n$. Here, the index $I$ takes values $I=\bar 1,\bar 2,\cdots,\bar n, 1,2,\cdots,n$, thereby labeling the $2n$ columns of the matrix $C$. These minors are anti-symmetric in its entries $I_1,\cdots\hskip -1pt,I_n$. Under a GL$(n)$ transformation, $C\mapsto R\cdot C$, the minors transform as 
\be \label{equ:minors-GL1}
(I_1\cdots I_n)\mapsto r\hs (I_1\cdots I_n)\,,
\ee  
where $r \equiv \det(R)$. The minors are therefore invariant under SL$(n)$ transformations, while ratios of minors are invariant under the whole GL$(n)$. The integral \eqref{equ:grassmanian-integral} is then invariant under a GL$(n)$ transformation as long as the 
integrand of \eqref{equ:grassmanian-integral} depends only on minors $A_n(C)=A_n((I_1\cdots I_n))$,
and scales as
\be \label{equ:An-GL1}
A_n(r\hs (I_1\cdots I_n))=r^{-(n-3)}\hs A_n((I_1\cdots I_n))\,.
\ee 
This scaling precisely cancels the scaling of the measure $\ud C$ and the $\delta$-function in~\eqref{equ:grassmanian-integral}. 

\subsubsection*{Little group covariance}

Next, we will explain how little group covariance constrains the functional form of the integrand $A_n$ in~\eqref{equ:grassmanian-integral}. 
Written in terms of the matrix $\Lambda$, the little group transformation is  
\beq
\Lambda \mapsto \rho \cdot \Lambda\,, \quad {\rm where} \quad \rho \equiv \text{diag}\left(\rho_1,\,\cdots,\rho_n,\frac{1}{\rho_1},\,\cdots,\frac{1}{\rho_n}\right) .
 \label{equ:lilgroup}
\eeq
In order for the inner product $C \cdot \Lambda$ to stay invariant, 
we  simultaneously perform the change of variables
\be \label{equ:lilgroup-Crho-1}
C\mapsto C\cdot \rho^{-1}\,.
\ee 
Since the integration measure in \eqref{equ:measure} is also invariant, the little group transformation only affects the minors $(I_1\cdots I_n)$ on which the function $A_n$ depends. 

\vskip 4pt
To understand how a minor transforms under the little group, notice that a barred column $\bar \imath$ in a minor scales as $1/\rho_{i}$ under \eqref{equ:lilgroup}, while an unbarred column $j$ scales as $\rho_j$. 
Hence, we obtain 
\be \label{equ:littlegroup-minors}
(\bar \imath_1\hs \bar \imath_2\hs\cdots \bar \imath_r \hs j_{r+1}\hs \cdots j_n)\mapsto \frac{1}{\rho_{i_1}}\cdots \frac{1}{\rho_{i_r}}\hs \rho_{j_{r+1}}\cdots \rho_{j_n}\hs  (\bar \imath_1\hs \bar \imath_2\hs \cdots \bar \imath_r\hs j_{r+1}\hs \cdots j_n)\,.
\ee 
Thus, the transformation properties of the minors are manifest in this notation: the columns $\bar \imath$ and $j$ transform exactly like the spinors $\bar\lambda_i$ and $\lambda_j$, respectively.

\vskip 4pt
Since the wavefunction coefficients scale under the little group as \eqref{equ:littlegroup-psin}, the function $A_n$ in~\eqref{equ:grassmanian-integral} must transform under \eqref{equ:littlegroup-minors} as
\be \label{equ:An-littlegroup}
A_n((I_1\cdots I_n))\mapsto \left(\prod_{i=1}^n\rho_i^{-2 h_i}\right)A_n((I_1\cdots I_n))\,.
\ee 
Together with the GL$(1)$ scaling in \eqref{equ:An-GL1}, this little group covariance constrains the function $A_n$.

\subsubsection*{Left and right branches}

It turns out that the orthogonal Grassmannian OGr$(n,2n)$ has two disconnected components that we refer to as the ``left" and ``right" branches.\footnote{Identifying the orthogonal Grassmannian OGr$(n,2n)$ with the space of (projective) pure spinors in $2n$ dimensions~\cite{Berkovits:2004bw}, these two components are defined by the chirality of the pure spinor. In fact, pure spinors are defined by certain quadratic constraints which imply that they must be Weyl spinors. Thus, there are two disconnected branches corresponding to left or right pure spinors.} Essentially, these are two different solutions to the quadratic orthogonality constraint \eqref{equ:CQC2}. Indeed, this constraint implies several quadratic constraints in terms of minors, with two inequivalent solutions (see Theorem~2.1 of~\cite{maazouz2025}). For instance, one of these constraints is
\be 
C\cdot Q\cdot C^T=0 \quad \Rightarrow \quad (\bar 1\bar 2\cdots \bar n)\hs ( 1\bar 2\cdots \bar n)=0\,,
\ee 
which can be solved by setting either $(1\bar 2\cdots\bar n)$ or $(\bar 1\bar 2\cdots\bar n)$ to zero. These two solutions correspond to the two different branches. We take the right branch to be defined by $(1\bar 2\cdots\bar n)\equiv0$, and the left branch by $(\bar1\bar 2\cdots\bar n) \equiv 0$.\footnote{More generally, the solutions to the orthogonality constraint on $C$ must satisfy 
	\be \label{equ:leftright}
	(I_1\cdots I_n)=\pm\hs\frac{(-1)^n}{n!}\hs Q_{I_1J_1}\cdots Q_{I_nJ_n} \epsilon^{J_1\cdots J_nK_1\cdots K_n}(K_1\cdots K_n)\,,
	\ee
	where the overall sign determines whether we are in the right ($+$) or left ($-$) branch. In this equation, we are implicitly summing over the indices $J_i$ and $K_i$, and we chose the Levi-Civita tensor in $2n$ dimensions with the convention $\epsilon^{\bar1\bar2\cdots\bar n12\cdots n}=1$. Indeed, we cannot go continuously from one branch to the other unless all minors vanish, and thus these branches are disconnected.}

\subsection{Gauge Fixing and Twistors}
\label{ssec:twistors}

To explicitly evaluate the Grassmannian integral~\eqref{equ:grassmanian-integral}, we fix the 
 GL$(n)$ redundancy by an appropriate gauge choice. The integral decomposes into two disconnected domains, corresponding to the left and right branches of OGr$(n,2n)$. For concreteness, we focus on a parameterization of the right branch. The results for the left branch could simply be obtained by interchanging barred and unbarred spinors. We will also show that the gauge-fixed integral has a natural origin in twistor space.

\subsubsection*{Gauge fixing}

To fix the GL$(n)$ redundancy in the right branch, we set the first $n$ columns of $C_{aI}$ (i.e.~the columns labelled by $\bar1,\bar 2,\cdots \hskip -1pt, \bar n$) to be the identity:
\be\label{equ:C-slice}
C =\begin{pmatrix} 1_{n\times n}\hs ,\hs  C_n\end{pmatrix} ,
\ee 
where $C_n$ is a $n\times n$ matrix with elements $(C_n)_{ij} \equiv -c_{ij}$.  
Note that the constraint $C\cdot Q\cdot C^T=0$ implies that the matrix $C_n$ must be anti-symmetric
\be 
0=C\cdot Q\cdot C^T=C_n+C_n^T \iff c_{ji} = - c_{ij}\,.
\ee 
Thus, the right branch of OGr$(n,2n)$ can be parameterized by the $n(n-1)/2$ independent parameters $c_{ij}$ and we can write the integral \eqref{equ:grassmanian-integral} as
\be 
\label{equ:tpsi-int}
\psi_n=\int \ud c_{ij} \, \delta\big(\lambda_i^\alpha- c_{ij} \bar\lambda_j^\alpha\big)\,A_n(c_{ij})\,,
\ee 
where $c_{ij} \bar\lambda_j^\alpha$ includes an implicit sum over the index $j=1,2,\cdots\hskip -1pt,n$, and the anti-symmetry of $c_{ij}$ trivializes the orthogonality constraint in the measure \eqref{equ:measure}. 
The integrand $A_n(c_{ij})$ is given by the function $A_n((I_1\cdots I_n))$ evaluated for the gauge-fixed matrix \eqref{equ:C-slice}.

\subsubsection*{Little group covariance}

It is easy to see from \eqref{equ:tpsi-int} that the parameters $c_{ij}$ transform under the little group as $c_{ij}\mapsto \rho_i\hs \rho_j\hs c_{ij}$, i.e.~in the same way as the angle bracket~$\langle ij\rangle$. However, the minors of the gauge-fixed matrix \eqref{equ:C-slice}, and hence the functions $A_n(c_{ij})$, do {\it not} transform as expected from \eqref{equ:An-littlegroup} under the little group.
The reason is that the overall scale of the minors has been fixed in this gauge by setting $(\bar 1\bar2\cdots \bar n)=1$, which spoils their little group covariance. To recover the transformation law \eqref{equ:An-littlegroup}, we restore this overall scale by adding the appropriate factors of $(\bar 1\bar2\cdots \bar n)$, leaving it unfixed.\footnote{Essentially, this makes the minors projective, thus restoring the GL$(1)$ subgroup of GL$(n)$ given by overall rescalings of the matrix $C$.} By rescaling the minors of the gauge-fixed matrix as in~\eqref{equ:minors-GL1}, with $r=(\bar 1\bar2\cdots \bar n)$, and using \eqref{equ:An-GL1}, we  get
	\be \label{equ:An-to-Ancij}
	A_n((I_1\cdots I_n))=
	(\bar 1\bar 2\cdots \bar n)^{-(n-3)}\hs A_n(c_{ij}) \,.
	\ee 
	While $A_n((I_1\cdots I_n))$ now transforms as \eqref{equ:An-littlegroup} under the little group, the function $A_n(c_{ij})$ transforms as
	\be 
	A_n(c_{ij})\mapsto \left(\prod_{i=1}^n \rho_i^{-2h_i -(n-3)}\right)A_n(c_{ij})\,.
	\label{equ:An-LG-fixed}
	\ee 
	We will use this when bootstrapping the Grassmannian correlators in a fixed gauge.

\subsubsection*{Grassmannian from twistors}

We will now take the opportunity to show how the cosmological Grassmannian also emerges via a different route, namely through the use of twistors.   This makes contact with the previous work~\cite{Baumann:2024ttn}, where twistors were used to describe the three-point correlators of conserved currents in three-dimensional conformal field theories.

\vskip 4pt
In Appendix~\ref{app:twistors}, we give a brief introduction to twistors and their recent application to correlators of conserved currents. There we show that an $n$-point correlator in (dual) twistor space can be written as
\begin{equation}
	\label{eq:twistorcij}
	F_n(W_i \cdot W_j) = \int \ud c_{ij} \: A_n(c_{ij}) \: \exp \left( -\frac{i}{2} c_{ij} W_i \cdot W_j \right) ,
\end{equation}
where the dual twistor associated to the particle $i$ has components $W_{i,A}=(\bar\mu_{i,\alpha},\bar\lambda_i^\beta)$. The inner products are defined as
$W_i \cdot W_j\equiv W_{i,A}\Omega^{AB}W_{j,B}$, with the symplectic form $\Omega^{AB}$ introduced in Appendix~\ref{app:twistors}. The function $A_n(c_{ij})$ is the correlator written in terms of the Schwinger parameters~$c_{ij}$. In \cite{Baumann:2024ttn},  this function was bootstrapped for $n=3$.

\vskip 4pt
Given the correlator $F_n(W_i \cdot W_j)$, we can obtain the momentum-space correlator by performing the following \textit{half Fourier transform}:
\begin{equation}
	\tilde F_n(\lambda_i, \bar \lambda_i) = \int 
	\ud^2 \bar\mu_k \, \exp(-i \lambda_k \cdot \bar\mu_{k}) \, F_n(W_i \cdot W_j)\,.
\end{equation}
Substituting \eqref{eq:twistorcij}, and integrating over $\bar\mu_k$, we get
\begin{equation}
	\tilde F_n(\lambda_i, \bar \lambda_i) = \int \ud c_{ij} \: 
	 \delta(\lambda_k^\alpha - c_{kl} \bar \lambda_l^\alpha)\, A_n(c_{ij}) \,,
	\label{eq:grassfromtwist}
\end{equation}
where we have ignored an overall numerical factor.
This is exactly \eqref{equ:tpsi-int}, which captures the right branch of the orthogonal Grassmannian.  
We would get the equivalent result for the left branch of the orthogonal Grassmannian 
if we replace an odd number of dual twistors $W_i$ with twistors $Z_i$ in \eqref{eq:twistorcij} and then perform the half Fourier transform. Different choices of (dual) twistors amount to different charts of the two branches.

\vskip 4pt
In~\cite{Baumann:2024ttn}, we emphasized that twistors provide natural variables in which both conformal symmetry and current conservation are manifest. Remarkably, the Grassmannian formalism developed here shares this advantage by simultaneously trivializing all isometries and enforcing current conservation. Furthermore, as we will see in Section~\ref{ssec:factorization}, Grassmannian correlators have another appealing feature: their analytic structure is governed by the elegant factorization property~\eqref{equ:Res-S=0-intro}, which we will exploit to construct four-point correlators directly in Grassmannian space.

\subsection{Grassmannian Integrals}

We now discuss how to compute the gauge-fixed integral~\eqref{equ:tpsi-int}. Although we will focus on the right branch, the same can be done for the left branch by choosing a different parameterization. 

\vskip 4pt
To begin with, we note that the $2n$-dimensional $\delta$-function, 
\be \label{equ:delta-l-cij-lb}
\delta\big(\lambda_i^\alpha- c_{ij} \bar\lambda_j^\alpha\big)\,,
\ee 
includes three-momentum conservation. 
Hence, $3$ out of the $2n$ constraints yield the $\delta$-function of momentum conservation. Only the remaining $2n-3$ equations will constrain the $n(n-1)/2$ parameters $c_{ij}$. Since these equations are linear in $c_{ij}$, they will force them to live in a solution space of dimension $(n-2)(n-3)/2$. 
Moreover, linearity ensures that a solution can always be parameterized as
\be \label{equ:cij-taugamma}
c_{ij}(\tau_\gamma)\equiv c_{ij}^{\star}+ \sum_\gamma d_{ij}^{\hs \gamma}\hs\tau_\gamma\,, \quad {\rm with} \quad c_{ij}^\star=\frac{\langle ij\rangle}{E}\,,
\ee 
where $c_{ij}^\star$ is a particular solution to \eqref{equ:delta-l-cij-lb} and $\tau_\gamma$
are $(n-2)(n-3)/2$ independent parameters. Following Section~2.1 of \cite{Arkani-Hamed:2009ljj}, this implies that \eqref{equ:delta-l-cij-lb} can be expressed as
\be \label{equ:linear-cons-n}
\delta\big(\lambda_i^\alpha- c_{ij} \bar\lambda_j^\alpha\big) =\delta\big(\vec k_1 + \cdots + \vec k_n \big)\times J \hs \int \ud \tau_\gamma \, \delta\big(c_{ij} -c_{ij}(\tau_\gamma)\big)\,,
\ee 
where $J$ is a Jacobian factor that depends on the choice of the parameters $\tau_\gamma$.
Here, we see explicitly how the $2n$-dimensional $\delta$-function \eqref{equ:delta-l-cij-lb} splits into the three-dimensional delta function $\delta(\vec k_1 + \cdots + \vec k_n)$ enforcing momentum conservation, and the remaining $2n-3$ delta functions $\int \ud\tau_\gamma\,\delta(c_{ij} -c_{ij}(\tau_\gamma))$. The latter constrain the  parameters $c_{ij}$ to have the form $c_{ij}=c_{ij}(\tau_\gamma)$ in \eqref{equ:cij-taugamma} for some values of the parameters~$\tau_\gamma$. 

\vskip 4pt
The correlator in \eqref{equ:tpsi-int} can then be written as 
	\be 
	\label{equ:tpsi-int2}
	 \psi_n^{\hs \prime} =  J\hs \int \ud \tau_\gamma \, A_n(c_{ij}(\tau_\gamma))\,,
	\ee 
	where the prime indicates that we have dropped the overall delta function $\delta\big(\vec k_1 + \cdots + \vec k_n \big)$. For tree-level wavefunction coefficients of massless spinning particles, $A_n$ will be a rational function of minors, which in turn are homogeneous polynomials of $c_{ij}$.  The integrand $A_n$ is therefore a rational function of the parameters $\tau_\gamma$, so the integral \eqref{equ:tpsi-int2} can be evaluated by summing the residues at the poles of $A_n$. Which residues are included in the sum depends on the choice of integration contour.

\subsubsection*{$\boldsymbol{n=3}$}

 In the case $n=3$, there is only one allowed solution for $c_{ij}$, which is $c_{ij}=\langle ij\rangle/E$. Equation~\eqref{equ:linear-cons-n} then becomes 
\be \label{equ:linear-cons-n=3}
\delta\big(\lambda_i^\alpha- c_{ij} \bar\lambda_j^\alpha\big) =\frac{1}{4}\hs \delta\big(\vec k_1 + \vec k_2+ \vec k_3 \big)\hs \delta\Big(c_{ij} -\frac{\langle ij\rangle}{E}\Big)\,,
\ee 
where no integrals are left because the number of remaining free parameters $\tau_\gamma$ is $(n-2)(n-3)/2=0$ for $n=3$. The precise Jacobian $J=1/4$ has been computed in Appendix C of \cite{Baumann:2024ttn}. This implies that the stripped wavefunction coefficient is
\be \label{equ:psi3tilde-A3cij}
 \psi_3^{\hs \prime} = \frac{1}{4}\hs
  A_3\left(c_{ij}\right)\bigg|_{c_{ij}=\langle ij\rangle/E}\,.
\ee 
Note that this correlator is supported only in the right branch because we used a parameterization that only covers this branch. To describe correlators supported in the left branch, we need the analog formulae for a different parameterization that covers this branch. One choice is the gauge where the last three columns of the matrix $C$ are the identity: 
	\be \label{equ:C3-Left-1}
	C = \left(\begin{matrix}
		0&-c_{12}&-c_{13}&1&0&0\\
		c_{12}&0&-c_{23}&0&1&0\\
		c_{13}&c_{23}&0&0&0&1
	\end{matrix}\right) .
	\ee 
In this parameterization, we find
\be \label{equ:linear-cons-n=3-left}
\delta\big( \bar\lambda_i^\alpha- c_{ij}\lambda_j^\alpha\big) =\frac{1}{4}\hs \delta\big(\vec k_1 + \cdots + \vec k_n \big)\hs \delta\Big(c_{ij} +\frac{\langle  \bar \imath \bar \jmath\rangle}{E}\Big)\,,
\ee 
and thus the stripped wavefunction is
\be \label{equ:psi3tilde-A3cijX}
\psi_3^{\hs \prime} = \frac{1}{4}\hs 
A_3 \left(c_{ij}\right)\bigg|_{c_{ij}=-\langle \bar \imath \bar \jmath \rangle/E}\,.
\ee 
In Section~\ref{sec:3pt}, we will derive explicit formulae for three-point functions $A_3$ in both the right and left branches, but for now let us simply stress that the integral \eqref{equ:grassmanian-integral} over the Grassmannian is trivial due to the $\delta$-functions~\eqref{equ:linear-cons-n=3} and  \eqref{equ:linear-cons-n=3-left}.

\subsubsection*{$\boldsymbol{n=4}$}

 In the case $n=4$, there is exactly $(n-2)(n-3)/2=1$ free parameter $\tau$  in \eqref{equ:cij-taugamma}. We can choose this parameterization  to be 
\be \label{equ:cijtau}
c_{ij}(\tau)=\frac{\langle ij\rangle}{E}+\tau\hs \frac{\epsilon_{ijkl}\hs\langle \bar k \bar l\rangle}{2}\,,
\ee 
Ignoring the Jacobian, which is just an overall numerical factor, we can write the stripped wavefunction as the following integral: 
\be \label{equ:tpsi-int3}
 \psi_4^{\hs \prime} =\int \frac{\ud\tau}{2\pi i} \,A_4(c_{ij}(\tau))\,,
\ee 
where the factor of $2\pi i$ was introduced for future convenience. 
In Section~\ref{sec:4pt}, we will explain how to bootstrap the integrand $A_4$ and present the case of four scalars exchanging a photon or a graviton. In Section~\ref{sec:YM}, we derive the four-point correlator in Yang--Mills theory.

\section{Three-Point Functions}
\label{sec:3pt}

In this section, we will use little group covariance to bootstrap the three-point functions for arbitrary spins (as in~\cite{Baumann:2024ttn}). Since there are many different, but equivalent, ways of expressing results in terms of minors, it is convenient to first derive the three-point functions in a fixed gauge and then introduce a natural uplift to gauge-invariant expressions in terms of minors.

\subsection{Gauge-Fixed Results}
For concreteness, we focus on correlators with all-plus helicities $\psi_{3,+++}$. We first consider the right branch and choose the gauge \eqref{equ:C-slice} for the Grassmannian matrix,
\be 
C =\left(\begin{matrix}
	1&0&0&0&-c_{12}&-c_{13}\\
	0&1&0&c_{12}&0&-c_{23}\\
	0&0&1&c_{13}&c_{23}&0
\end{matrix}\right) ,
\label{equ:C3-R}
\ee 
where the parameters $c_{ij}$ scale under the little group as $c_{ij}\mapsto \rho_i\hs \rho_j\hs c_{ij}$. In this parameterization, a generic three-point function may be obtained by a power-law ansatz
\be \label{equ:A3-right}
 A_{3,+++}^{(\rm{leading})}(c_{ij})=c_{12}^{-n_3}\hs c_{23}^{-n_1}\hs c_{31}^{-n_2}\,,
\ee 
where the superscript will be explained later.  
Imposing the expected little group covariance~\eqref{equ:An-LG-fixed}, for all-plus helicities, the exponents $n_i$ become
\be \label{equ:ni-exponents}
n_i=\ell_j+\ell_k-\ell_i\,,
\ee 
where $\{i,j,k\}$ is a cyclic permutation of $\{1,2,3\}$. 

\vskip 4pt
Next, we turn to the left branch and choose the parameterization \eqref{equ:C3-Left-1} for
 the matrix $C$, where the parameters $c_{ij}$ scale under the little group as $c_{ij}\mapsto \rho_i^{-1}\hs \rho_j^{-1}\hs c_{ij}$. 
In this parameterization, a generic three-point function, with all-plus helicities, is given by 
\be \label{equ:A3-left}
A_{3,+++}^{(\rm{higher})}(c_{ij})=c_{12}^{n_3}\hs c_{23}^{n_1}\hs c_{31}^{n_2} \,,
\ee 
where the exponents $n_i$ are the same as in \eqref{equ:ni-exponents}.

\vskip 4pt
When the particles have equal spins $\ell_i = \ell$, the above results become
\be\label{equ:A3cij}
\begin{aligned}
A_{3,+++}^{(\rm{leading})}(c_{ij}) &=  \left( c_{12}\hs c_{23}\hs c_{31} \right)^{-\ell}\,,\\
A_{3,+++}^{(\rm{higher})}(c_{ij}) &= \left( c_{12}\hs c_{23}\hs c_{31} \right)^\ell \,.
\end{aligned}
\ee 
As some of us proved in~\cite{Baumann:2024ttn}, the all-plus three-point correlator \eqref{equ:psi3tilde-A3cij}  obtained from $A_{3,+++}^{(\rm{leading})}$ corresponds to the {\it leading} interactions in the bulk (i.e.~YM or GR when the particles have spin $\ell=1$ or $2$, respectively), while the correlator obtained from $A_{3,+++}^{(\rm{higher})}$ 
corresponds to {\it higher-derivative} interactions in the bulk (i.e.~$F^3$ or $R^3$ when the particles have spin $\ell=1$ or $2$, respectively).\footnote{To be more precise, for the case of leading interactions, the resulting correlator \eqref{equ:psi3tilde-A3cij} is  the {\it discontinuity} of the momentum-space correlator with respect to $k_1^2,\hs k_2^2,\hs k_3^2$, which solves the homogeneous SCT constraint~\eqref{equ:sct-hom}; see \cite{Baumann:2024ttn} for more details.} This explains the superscripts in \eqref{equ:A3-right} and~\eqref{equ:A3-left}.
Remarkably, these two ans\"atze yielded precisely the two correlators corresponding to these different bulk interactions. 

\subsection{Gauge-Invariant Results}
\label{sec:gaugeinv3pt}

We will now show how to write these three-point correlators in terms of minors of the Grassmannian. This way of writing them does not depend on the choice of parameterization of each branch. 

\vskip 4pt
   To begin with, it is useful to identify  little group-invariant building blocks. Since there are no little group-invariant minors for $n=3$, these building blocks must involve products of two minors. Indeed, a product of two minors that is manifestly little group invariant is
   \be \label{equ:E3}
   {\cal K}\equiv (i \bar \imath j )\hs (\bar\jmath k \bar k) \,,
   \ee 
where $\{i,j,k\}$ is a cyclic permutation of $\{1,2,3\}$. 
It turns out that this is the only independent little group-invariant product of minors at three points.\footnote{A simple way to show this is by going to a specific gauge, such as \eqref{equ:C3-R} in the right branch. It can then be seen by inspection that, ignoring signs, all non-zero minors in the right branch are given by 
either $1$ or $c_{ij}$ or $c_{ij}c_{jk}$, for some $i,j,k$. Hence, the product of $r$ minors (or $r+r'$ minors divided by $r'$ minors) will be given in this parameterization by $(\bar 1\bar 2\bar3)^r\hs  c_{12}^{m_3}\hs c_{23}^{m_1}\hs c_{31}^{m_2}$, where we restored the overall scale $(\bar 1\bar 2\bar3)$ to make the little group scaling of the minors manifest, as explained above equation \eqref{equ:An-to-Ancij}. Since the parameters $c_{ij}$ scale under the little group as~$\langle ij\rangle$, this product will only be little group invariant if $m_1=m_2=m_3= r/2$, which implies that $r=2m$ must be even. Hence, this product must be equal to ${\cal K}^m$, where ${\cal K}= c_{12}\hs c_{23}\hs c_{31}$ is precisely \eqref{equ:E3} in the gauge~\eqref{equ:C3-R} after setting again $(\bar 1\bar 2\bar3)=1$. Similarly, in the gauge \eqref{equ:C3-Left-1}, parameterizing the left branch, the only little group-invariant product of minors is ${\cal K}^m$, with ${\cal K}= -c_{12}\hs c_{23}\hs c_{31}$, which equals \eqref{equ:E3} in this gauge.  \label{footnote-c12c23c31}} 
All other little group-invariant products of minors either vanish or are given in terms of ${\cal K}$,~e.g.~$(i j \bar \jmath)\hs (j \bar \imath \bar \jmath) =\pm{\cal K}$, where the sign depends on the branch. 

\vskip 4pt
Equipped with the little group-invariant building block ${\cal K}$, we will uplift \eqref{equ:A3-right} and~\eqref{equ:A3-left} to results in terms of minors by imposing that only ${\cal K}$ appears in the denominator, and the order of its singularity is as small as possible. Assuming, without loss of generality, that $\ell_1\leq \ell_2\leq \ell_3$, we can write \eqref{equ:A3-right}  as
\begin{align} \label{equ:a3leading-ppp}
A_{3,+++}^{(\text{leading})}((I_1I_2I_3)) &=\frac{(\bar 1\bar 2\bar 3)^{2\ell_1}\hs (\bar 1\bar 2 1)^{2(\ell_2-\ell_1)}(\bar 1\bar 31)^{2(\ell_3-\ell_1)}}{{\cal K}^{\ell_2+\ell_3-\ell_1}}\,, \\[6pt]
 \label{equ:a3higher-ppp}
A_{3,+++}^{(\text{higher})}((I_1I_2I_3)) &=\frac{(\bar 1\bar 2 3)^{2\ell_1}\hs (\bar 1\bar 21)^{2(\ell_2-\ell_1)}\hs (\bar 3\bar11)^{2(\ell_3+\ell_1)}}{{\cal K}^{\ell_1+\ell_2+\ell_3}}\,,
\end{align} 
where all powers of the minors in the numerator are non-negative.\footnote{Since these minors satisfy several identities from the so-called Pl\"ucker relations and the orthogonality condition~\cite{maazouz2025}, there are many equivalent ways of writing the numerators as a product of minors.} 
These results are manifestly little group covariant, satisfy GL$(1)$ invariance, and the power of ${\cal K}$ in the denominator is as small as it can be.  For equal spins $\ell_i=\ell$, the above results reduce to
\begin{align} 
A_{3,+++}^{({\rm leading})} &= \left(\frac{(\bar 1\bar 2\bar 3)^{2}}{{\cal K}}\right)^\ell \, , \\[6pt]
A_{3,+++}^{(\text{higher})} &=\left(\frac{(\bar 1\bar 2 3)^{2}\hs  (\bar 3\bar11)^{4}}{{\cal K}^{3}}\right)^\ell =\left(\frac{(\bar 1\bar 2 3)^{2}\hs(\bar 1 2\bar 3)^{2}\hs( 1\bar 2 \bar3)^{2}}{{\cal K}^{3}}\right)^\ell\,, \label{equ:higher}
\end{align}
where we used some minor identities in the last equality.

\vskip 4pt
 Equation~\eqref{equ:a3leading-ppp} trivially reduces to \eqref{equ:A3-right} in the gauge~\eqref{equ:C3-R}, where ${\cal K}=c_{12}c_{23}c_{31}$, $(\bar 1\bar 2 1)= c_{13}$ and $(\bar 1\bar 3 1)= -c_{12}$. Furthermore, if $\ell_1>0$ (i.e.~if there are no scalars), it is manifest in this expression that the correlator is only supported in the right branch because the minor $(\bar1\bar2\bar3)$ vanishes in the left branch.  Similarly, equation~\eqref{equ:a3higher-ppp} reduces trivially to \eqref{equ:A3-left} in the gauge \eqref{equ:C3-Left-1}. Moreover, if there are no scalars (i.e., if $\ell_1>0$), the correlator is supported only on the left branch, since $(\bar 1\bar 23)$ vanishes on the right branch.

\vskip 4pt
We can get the corresponding correlators with different helicity configurations by interchanging barred and unbarred columns $i\leftrightarrow \bar \imath$. For instance, to get the $++-$ correlator from the $+++$ correlators \eqref{equ:a3leading-ppp} and \eqref{equ:a3higher-ppp}, we flip the columns $3\leftrightarrow\bar 3$. This yields 
\begin{align} \label{equ:a3leading-ppm}
A_{3,++-}^{(\text{leading})} &=\frac{(\bar 1\bar 2 3)^{2\ell_1}\hs (\bar 1\bar 2 1)^{2(\ell_2-\ell_1)}(\bar 1 31)^{2(\ell_3-\ell_1)}}{{\cal K}^{\ell_2+\ell_3-\ell_1}} \xrightarrow{\ \ell_i =\ell\ } \left(\frac{(\bar 1\bar 2 3)^{2}}{\cal K}\right)^\ell\,, \\[6pt]
 \label{equ:a3higher-ppm}
A_{3,++-}^{(\text{higher})} &=\frac{(\bar 1\bar 2 \bar3)^{2\ell_1}\hs (\bar 1\bar 21)^{2(\ell_2-\ell_1)}\hs ( 3\bar11)^{2(\ell_3+\ell_1)}}{{\cal K}^{\ell_1+\ell_2+\ell_3}} \xrightarrow{\ \ell_i =\ell\ }  \left(\frac{(\bar 1\bar 2 \bar3)^{2}\hs(\bar 1 2 3)^{2}\hs( 1\bar 2 3)^{2}}{{\cal K}^{3}}\right)^\ell \,, 
\end{align} 
which have support in the left and right branches, respectively. In the last step in (\ref{equ:a3higher-ppm}), we used some minor identities. Analogously, we can get the correlators in any desired helicity configuration by applying the appropriate flips of barred and unbarred columns in the minors. 

\vskip4pt 
Let us stress that the main difference between the correlators corresponding to the leading and higher-derivative interactions is 
{\it not} the branch in which they are supported. In fact, this branch is not the same for different helicity configurations: For leading interactions, the $+++$ correlator \eqref{equ:a3leading-ppp} is supported in the right branch, while the $++-$ correlator \eqref{equ:a3leading-ppm} is supported in the left branch. Conversely, for higher-derivative interactions, the $+++$ and $++-$ correlators are supported in the left and right branches, respectively. 

\subsection{Flat-Space Limit}
\label{ssec:A3-flat-space}

The feature that really distinguishes between the leading and higher-derivative interactions at the level of the Grassmannian correlator $A_3$ is the order of the pole at ${\cal K} =0$.  As we will now show, it is possible to extract the corresponding flat-space amplitude $M_3$ in momentum space via a substitution rule performed on the minors that multiply this pole, while the order of the pole determines the mass dimension of $M_3$ and thus distinguishes between the contributions from leading and higher-derivative interactions. 
This substitution rule encapsulates a two-step process in a single step: it effectively computes the momentum-space correlator $\psi_3$ and then extracts the coefficient of its leading $E\to 0$ singularity.

\vskip 4pt
For clarity, let us work in the left branch, using the parameterization
\beq
C =\left(\begin{matrix}
	1&0&-c_{13}&0&-c_{12}&0\\
	0&1&-c_{23}&c_{12}&0&0\\
	0&0&0&c_{13}&c_{23}&1
\end{matrix}\right).
\eeq
The momentum-space correlator $\psi_3$ associated to $A_3$ can be obtained by the substitution
\beq
\mathcal{K} \mapsto E \, \frac{[23][31]}{[12]^3} \,, \quad {\rm with} \quad E \equiv k_1+k_2+k_3\,,
\eeq
where $[i j]$ are the flat-space square brackets. 
Furthermore, we can leverage the $\mathrm{GL}(1)$ invariance of the Grassmannian to rescale all minors by $([12]^3/[23][31])^{1/2}$, so that $\mathcal{K}$ evaluates exactly to~$E$. It is then straightforward to verify that the only minors that are not set to zero, in this limit, are those with exactly two barred columns, which evaluate to\footnote{Similarly, in the right branch, we obtain the flat-space amplitude by computing the leading singularity at ${\cal K}=0$ and setting only minors with exactly one barred column to be non-zero: 
	\beq
	( i j\bar k)^2 \mapsto \frac{\langle  i j\rangle^3}{\langle  j k\rangle \langle  k i\rangle}\quad {\rm and} \quad
	( i\bar \jmath j)^2  \mapsto \frac{\langle  i j\rangle \langle  k i\rangle}{\langle  j k\rangle}\,,
	\eeq
	where $\{i,j,k\}$ is some permutation of $\{1,2,3\}$ and we have neglected the subleading terms. While the flat-space limit of the left branch yields flat-space amplitudes in the anti-holomorphic configuration with only square brackets, the limit on the right branch yields amplitudes in the holomorphic configuration with only angle brackets.}
\beq
\label{equ:3pt-flat-square}
(\bar \imath \bar \jmath k)^2 \mapsto \frac{[ij]^3}{[jk][ki]} + \mathcal{O}(E) \quad {\rm and} \quad
(\bar \imath \bar \jmath j)^2 \mapsto \frac{[ij][ki]}{[jk]} + \mathcal{O}(E)\,,
\eeq 
where $\{i,j,k\}$ is some cyclic permutation of $\{1,2,3\}$. 

\vskip 4pt
We now apply these substitution rules to specific Grassmannian three-point functions $A_4$
and verify that they reproduce the correct momentum-space amplitudes. For instance, the flat-space limit of the $++-$ correlator in \eqref{equ:a3leading-ppm} yields
\beq
\lim_{{\cal K} \to 0}A_{3,++-}^{({\rm leading})} \mapsto \frac{1}{E^{\ell}}\left(\frac{[12]^3}{[23][31]}\right)^\ell \equiv \frac{1}{E^{\ell}} M_{3,++-}^{({\rm leading})}\,.
\eeq 
We see that the coefficient of the singularity is indeed the known three-point amplitude with helicities $++-$. We have chosen the helicity configuration $++-$ because its corresponding flat-space amplitude is non-trivial for leading interactions like YM or GR, while $M_{3,+++}^{({\rm leading})}$ vanishes. Meanwhile, the flat-space limit of the $+++$ correlator in \eqref{equ:higher} gives the known three-point amplitude for higher-derivative interactions:
\beq 
\lim_{{\cal K} \to 0} A_{3,+++}^{({\rm higher})} \mapsto \frac{1}{E^{3\hs\ell}} \left([12][23][31]\right)^\ell = \frac{1}{E^{3\hs\ell}} M_{3,+++}^{({\rm higher})}\, .
\eeq
Here, we can clearly see that the resulting flat-space amplitudes $M_{3,++-}$ and $M_{3,+++}$ have different mass dimensions due to the different orders of the pole at ${\cal K}=0$.

\section{Four-Point Functions}
\label{sec:4pt}

In the previous section, we showed that the three-point functions in Grassmannian space can be bootstrapped from little group covariance alone. To derive higher-point functions, however, we need additional input beyond just kinematics. Recall from equation \eqref{equ:tpsi-int3} that four-point wavefunction coefficients are given by the following integral 
\be 
\label{equ:tpsi-int4}
\psi_4^{\hs \prime} =  \int \frac{\ud \tau}{2\pi i} \,A_4(c_{ij}(\tau))\,.
\ee 
In this section, we will show that the function $A_4$ is determined by its factorization limits, which follow from 
unitarity.  
Remarkably, this factorization property for $A_4$ is the same as for scattering amplitudes. We also show how the flat-space limit is implemented at the level of the Grassmannian correlators and that it leads to the expected total energy singularities in momentum space. We illustrate the bootstrap procedure for the case of photon and graviton exchange, as well as for Yang--Mills theory (in Section~\ref{sec:YM}).

\subsection{Mandelstams}
\label{ssec:Mandelstams}

At four points, it is useful to define the following
 little group-invariant minors 
\beq
\begin{aligned}
	S&\equiv  (\bar 1\bar 212 )\,,\\
	T &\equiv (\bar 1\bar 414 )\,,\\
	U &\equiv (\bar 1\bar 313 )\,.
\end{aligned}
\eeq
For reasons that will become clear below,
 we will refer to these minors as ``Mandelstams". 
  It turns out that any little group-invariant product of minors at four points can be written in terms of $S,T,U$, and thus they are the basic building blocks for any little group-invariant function of minors.\footnote{One can easily show this in a specific gauge, such as \eqref{equ:C-slice}. It can be seen by inspection that any product of $r$ minors 
  	can be written as linear combinations of terms of the form~$(\bar 1\bar 2\bar3\bar4)^r c_{12}^{m_{12}}\hs c_{34}^{m_{34}} \hs c_{14}^{m_{14}}\hs c_{23}^{m_{23}}\hs c_{13}^{m_{13}}\hs c_{24}^{m_{24}}$. Here, we restored the overall scale $(\bar 1\bar 2\bar3\bar4)$ to make the little group scaling of the minors manifest, as explained above equation \eqref{equ:An-to-Ancij}. This will be little group invariant if $\sum_{j\neq i}m_{ij}=r$. Noting that $S+T-U= 2\hs c_{13}c_{24}$, and similarly for $S-T+U$ and $-S+T+U$, this term is of the form~$(-S+T+U)^{m_{12}}(S-T+U)^{m_{14}}(S+T-U)^{m_{13}}$ up to an overall constant after setting  $(\bar1\bar2\bar3\bar4)=1$, and thus it is always a function of the Mandelstams $S,T,U$.} 
In the gauge \eqref{equ:C-slice}, these Mandelstams become
\beq
\begin{aligned}
	S&= c_{13}c_{24}-c_{14}c_{23}\,,\\
	T &= c_{13}c_{24}-c_{12}c_{34}\,,\\
	U &=-c_{14}c_{23}-c_{12}c_{34}\,.
\end{aligned}
\eeq
Using
\eqref{equ:cijtau} for $c_{ij}(\tau)$, these expressions become quadratic polynomials of $\tau$. For instance, the variable $S(\tau)$ is
\be 
S(\tau)=\langle\bar1\bar2\rangle\langle \bar3\bar4\rangle\hs (\tau-\tau_s)\hs (\tau- \bar \tau_{s})\,,
\ee 
where we have defined
\begin{align}
\tau_s&\equiv \frac{1}{\langle\bar1\bar2\rangle\langle \bar3\bar4\rangle}\hs \frac{E_LE_R}{E}\,,\label{equ:tau-s}\\
\bar \tau_{s}&\equiv \frac{1}{\langle\bar1\bar2\rangle\langle \bar3\bar4\rangle}\hs \frac{\bar E_L\bar E_R}{E}\,,\label{equ:tau-sb}
\end{align}
with $k_s\equiv |\vec k_1+\vec k_2|$ and
\beq
\begin{aligned} 
E \equiv k_1+k_2+k_3+k_4 \qquad
E_L&\equiv k_1+k_2+k_s\,, \qquad \bar E_L \equiv k_1+k_2-k_s\,,\\
E_R&\equiv k_3+k_4+k_s\,, \qquad \bar E_R \equiv k_3+k_4-k_s\,.
\end{aligned}
\eeq
 Notice that, while the Mandelstam $S(\tau)$ is a rational function of the spinor helicity variables, the roots $\tau_s$ and $\bar \tau_{s}$ are not.  Lastly, it is easy to prove that the sum of the Mandelstams is 
\be 
{\cal E}(\tau)\equiv -\frac{(S+T+U)(\tau)}{2}= \tau\hs E\,.
\label{equ:cal-E}
\ee 
We see that this sum doesn't vanish, but is proportional to the total energy $E$.
Only in the limit $E \to 0$ does the relation $S+T+U=0$ hold, as expected for the standard Mandelstam variables in flat space.

\vskip 4pt
These Mandelstams will play a central role in bootstrapping the four-point Grassmannian correlators $A_4$. As we will see below, the analytic structure of $A_4$
 features poles both at $S+T+U=0$ and in the individual channels $S=0$, $T=0$ and/or $U=0$. These singularities encode information about the flat-space limit and on-shell factorization, respectively.

\subsection{Flat-Space Limit}
\label{ssec:flat-space}

In Section~\ref{ssec:A3-flat-space}, we showed that the three-point correlator $A_3$ has a pole when the little group-invariant structure ${\cal K}$ vanishes, and that the coefficient of that pole allows us to easily read off the corresponding momentum-space amplitude $M_3$. We will now show that the four-point correlator $A_4$ must have a pole at ${\cal E} = 0$, and that its coefficient is proportional to the corresponding momentum-space amplitude $M_4$. As we will see, this constraint is automatically satisfied for interactions constructed from three-point data.

\vskip 4pt
First, a pole at ${\cal E} = 0$ is required so that the integral \eqref{equ:main-integral} admits a singularity at $E = 0$, as discussed in Appendix~\ref{app:flat}. Indeed, any Grassmannian correlator $A_4$ contains a term of the form
\beq
A_4 \supset \frac{1}{{\cal E}^{r+1}} \mathcal{A}_4\,,
\eeq
where $r$ is a non-negative integer. We omit terms that are subleading in the $\mathcal{E} \to 0$ limit. The momentum-space correlator $\psi_4$ associated to $A_4$ can be obtained via a sum over $\tau$ residues. In particular, in Appendix~\ref{app:flat}, we show that only $\tau = 0$ is singular for $E \to 0$ and that the left branch does not contribute, so we can restrict the analysis to the study of this pole. In the insert below we show that the following substitution rule holds:
\beq
\label{equ:sub}
\lim_{{\cal E} \to 0} A_4 = \lim_{{\cal E} \to 0} \frac{1}{{\cal E}^{r+1}} \mathcal{A}_4((i j \bar k \bar l)) \mapsto \binom{2r}{r} \frac{1}{E^{2r+1}} \mathcal{A}_4(\langle i j \rangle [k l]) \equiv \frac{(2r)!}{E^{2r+1}} M_4\,,
\eeq
where we assumed that $\mathcal{A}_4$ only depends on minors $(i j \bar k \bar l)$ with two barred and two unbarred indices.\footnote{In the limit $E \to 0$, any other minor gives only a subleading contribution in momentum space. For some correlators, however, these other minors control the leading contribution in the flat-space limit. In such cases, one must keep their precise behavior rather than setting them to zero. Although none of the correlators studied in this paper exhibit this feature, the strategy outlined in Appendix~\ref{app:flat} yields the correct result even in those situations.}
Since the coefficient of the total energy singularity must be the known flat-space amplitude~$M_4$, this puts a constraint on the functional form of the leading singularity of the Grassmannian correlator, i.e.~on the function ${\cal A}_4$.

\vskip 6pt
\begin{eBox3}
{\bf Derivation}: \ 
We now prove the substitution rule (\ref{equ:sub}). 
Taking the residue of $A_4$ at $\tau = 0$  amounts to evaluating
\beq
\psi_4 \supset \frac{1}{E} \frac{1}{r!} \partial^r_{x} [\mathcal{A}_4((i j\bar k \bar l)(x))] \Big|_{x = 0} \,,
\eeq
where we defined the dimensionless variable $x \equiv \tau E$. In the  limit $E \to 0$, it is straightforward to verify from \eqref{equ:cijtau} that
\beq
(ij\bar k\bar l)(x) = \frac{\epsilon_{ijab}[ab][kl]}{2E^2} (x + R)^2 + \mathcal{O}\big(E^{-1}\big)
\,, \quad {\rm where} \quad R \equiv \frac{s}{[12][34]} \,.
\eeq
Crucially, all the expressions for $R$ that appear in each minor are equivalent to each other. 
Since $\mathcal{A}_4$ has weight $r$ in the minors, we  find
\begin{align}
	\psi_4 & \supset \frac{1}{E^{2r+1}} \mathcal{A}_4\bigg( \frac{\epsilon_{ijab}[ab][kl]}{2}\bigg) \, \frac{1}{r!} \partial^r_x[(x+R)^{2r}]\Big|_{x = 0} \nonumber \\[4pt]
	& \supset \binom{2r}{r} \frac{1}{E^{2r+1}} \mathcal{A}_4(\langle i j\rangle [kl]) \,,
\end{align}
where, in the second line, we have distributed the $r$ factors of $R$ among the arguments of $\mathcal{A}_4$ to restore the desired form. We have therefore obtained the result~\eqref{equ:sub}.
\end{eBox3}

\vspace{0.05cm}
\subsection{Factorization Limits}
\label{ssec:factorization}

Our main tool for bootstrapping the correlators for exchange diagrams will be the fact that unitarity imposes specific factorization properties on the results.

\vskip 4pt 
For concreteness, we consider the exchange of a massless particle in the $s$-channel. Unitarity then implies the following cutting rule in momentum space~\cite{Goodhew:2021oqg, Baumann:2021fxj}:
\be \label{equ:cutting}
\mathop{\mathrm{Disc}}_{k_s^2}[\psi_4^{\hs \prime}] = - \frac{1}{2k_s}\hs\sum_h \mathop{\mathrm{Disc}}_{k_s^2}[\psi_{3,L}^{\hs \prime \hs (-h)}]\,\mathop{\mathrm{Disc}}_{k_s^2}[\psi_{3,R}^{\hs \prime \hs (+h)}]\,,
\ee
where the discontinuity with respect to the internal energy $k_s$ is defined as 
\be\label{equ:Disc-def}
\mathop{\mathrm{Disc}}_{k_s^2}[\psi]= \psi(k_s)- \psi(-k_s)\,.
\ee 
This cutting rule has played an important role in constraining the space of consistent correlators in momentum space, especially when the correlators are rational functions. In Appendix~\ref{app:factorization}, we will show that 
\eqref{equ:cutting} implies that the function $A_4$ in \eqref{equ:tpsi-int4} has a simple pole at $S = 0$ (for the $s$-channel), where it factorizes as 
\be \label{equ:Res-S=0}
\mathop{\mathrm{Res}}_{S=0} A_4= \sum_h A_{3,L}^{(-h)} \hs A_{3,R}^{(+h)}\,,
\ee
where $A_{3,L}^{(-h)}$ and $A_{3,R}^{(+h)}$ are the relevant three-point correlators, and $h$ is the helicity of the exchanged particle. Remarkably, the result \eqref{equ:Res-S=0} is identical to the on-shell factorization of scattering amplitudes. 

\vskip 4pt
To check the factorization limit \eqref{equ:Res-S=0}, we will use the following identities for the minors of the four-point Grassmannian 
\be\label{equ:minors-factorized}
\begin{aligned}
(L_1L_2R_1R_2)&\ \xrightarrow{\ S=0\ }\ (L_1L_2 I_s)(R_1R_2\bar I_s)+(L_1L_2\bar I_s)(R_1R_2 I_s) \,, \\
(L_1L_2L_3R_1)&\ \xrightarrow{\ S=0\ }\	-(L_1L_2L_3)(R_1 I_s\bar I_s)\,,\\
(L_1R_1R_2R_3)&\ \xrightarrow{\ S=0\ }\	(L_1 I_s\bar I_s)(R_1R_2R_3)\,,
\end{aligned}
\ee	
where $L_i=\bar1,\bar2, 1,2, \: R_i=\bar3,\bar4, 3,4$, and $\bar I_s,I_s$ are the columns associated to the exchanged particle.  In the following, we will illustrate the bootstrap procedure in two instructive examples.

\subsubsection*{Photon exchange}

Consider the four-point function of conformally coupled scalars exchanging a photon in the $s$-channel. 
The factorization rule \eqref{equ:Res-S=0} states that the residue at $S=0$ must be given by
\be 
\mathop{\rm{Res}}_{S=0}A_{4,\gamma}= A_{3,L}^{(-1)}A_{3,R}^{(+1)}+A_{3,L}^{(+1)}A_{3,R}^{(-1)}\,.
\ee 
Using the result for $\langle O O^\dagger J^\pm \rangle$ derived in Section~\ref{sec:gaugeinv3pt}, we have 
\be \label{equ:3pt-OOJ}
\begin{aligned}
A_3^{(+1)}&=\frac{(1\bar 2\bar 3)(\bar 12\bar 3)}{{\cal K}}\,,\\
A_3^{(-1)}&=\frac{(1\bar 2 3)(\bar 12 3)}{{\cal K}}\,,
\end{aligned}
\ee
and the residue at $S=0$ is
\begin{align}\label{equ:res-s-Agamma}
\mathop{\rm{Res}}_{S=0}A_{4,\gamma} &= \frac{(1\bar 2\bar I_s)(\bar 12\bar I_s)}{(2 \bar I_s I_s)(I_s \bar 2\bar I_s)}\hs \frac{(3\bar 4 I_s)(\bar 34 I_s)}{(3 4 \bar 4)(4 \bar 3\bar 4)} + \frac{(1\bar 2 I_s)(\bar 12 I_s)}{(2 I_s \bar I_s)(I_s \bar 2\bar I_s)}\hs \frac{(3\bar 4 \bar I_s)(\bar 34\bar I_s)}{(3 \bar 4 4)(4 \bar 3\bar 4)}\nonumber\\[4pt]
&=\frac{(1\bar 2\bar 34)(\bar 123\bar 4)}{(2 34 \bar4)(\bar2\bar 3\bar 44)}\bigg|_{S=0}=\frac{(-S+T-U)^2}{(S+T+U)(-S+T-U)}\bigg|_{S=0}
\nonumber\\[4pt]
&=\frac{T-U}{S+T+U}\bigg|_{S=0}\,.
\end{align}
In the second line, we used \eqref{equ:minors-factorized} to combine the product of three-point minors into four-point minors.\footnote{Notice that the cross terms in the product vanish
\begin{align}
(1\bar 2\bar 34)(\bar 123\bar 4)\big|_{S=0} & =\left((1\bar 2 I_s)(\bar 34 \bar I_s)+(1\bar 2 \bar I_s)(\bar 34 I_s)\right) \left((\bar1 2 I_s)( 3\bar4 \bar I_s)+(\bar1 2 \bar I_s)( 3\bar4 I_s)\right) \nonumber \\[4pt]
& = (1 \bar 2 I_s) (\bar 1 2 I_s)(\bar 3 4 \bar I_s)(3 \bar 4 \bar I_s) + (1 \bar 2 \bar I_s)(\bar 1 2 \bar I_s)(\bar 3 4 I_s)(3 \bar 4 I_s) \,.
\end{align} 
For instance, the product $(1\bar 2\bar I_s)(\bar1 2 I_s)$ vanishes because the minors $(\bar1 2 I_s)$ and $(1\bar 2\bar I_s)$ vanish separately in the left and right branches, respectively.
} In the last line, we computed this residue on the support of $S=0$ to express the result in a convenient way.

\vskip 4pt
The factorization property~\eqref{equ:res-s-Agamma} determines the correlator up to contact terms that are singular only at $S+T+U=0$, and thus are regular at $S=0$. To fix these contact terms, we observe that we have just found
\beq
	A_{4,\gamma} = \frac{1}{S+T+U} \frac{T-U}{S} + (\mathrm{contact}) = -\frac{1}{S+T+U} \,P_1 \bigg( \frac{U-T}{S} \bigg) + (\mathrm{contact}),
\eeq
where $P_1$ is the first Legendre polynomial. This is the expected functional form of a spin-$1$ exchange without any additional contact term, and it is also compatible with the expectation of the flat-space limit
\be \label{equ:lead-E-Agamma}
\mathop{\rm{lim}}_{{\cal E}=0}A_{4,\gamma} = \mathop{\rm{lim}}_{{\cal E}=0}\frac{1}{{\cal E}}\frac{U-T}{2 S} \mapsto \frac{1}{E} \frac{u-t}{2 s} \,.
\ee 
This fixes the photon exchange correlator to be
\be \label{equ:A4-gamma}
\boxed{A_{4,\gamma} =\frac{1}{S+T+U}\frac{T-U}{S}}\ .
\ee 
We will show in Section~\ref{ssec:Int} that the $\tau$-integral \eqref{equ:tpsi-int4} of this Grassmannian correlator reproduces the known correlator in momentum space.

\subsubsection*{Graviton exchange}

Next, we will bootstrap the four-point function of scalars exchanging a graviton in the $s$-channel. As in the previous example, the residue at $S=0$ can be derived from on-shell factorization. Using the square of the three-point correlators~\eqref{equ:3pt-OOJ} for $\langle O O T^\pm \rangle$, the factorization property~\eqref{equ:Res-S=0} reads
\begin{align}\label{equ:res-s-Ag}
	\mathop{\rm{Res}}_{S=0}A_{4,g} &= \left(\frac{(1\bar 2\bar I_s)(\bar 12\bar I_s)}{(2 \bar I_s I_s)(I_s \bar 2\bar I_s)}\hs \frac{(3\bar 4 I_s)(\bar 34 I_s)}{(3 4 \bar 4)(4 \bar 3\bar 4)} \right)^2 + \left(\frac{(1\bar 2 I_s)(\bar 12 I_s)}{(2 I_s \bar I_s)(I_s \bar 2\bar I_s)}\hs \frac{(3\bar 4 \bar I_s)(\bar 34\bar I_s)}{(3 \bar 4 4)(4 \bar 3\bar 4)}\right)^2 \nonumber\\[4pt]
	&=\left(\frac{(1\bar 2\bar 34)(\bar 123\bar 4)}{(2 34 \bar4)(\bar2\bar 3\bar 44)} \right)^2\bigg|_{S=0} = \left(\frac{(-S+T-U)^2}{(S+T+U)(-S+T-U)}\right)^2 \bigg|_{S=0} 
	\nonumber\\[4pt]
	&=\left(\frac{T-U}{S+T+U}\right)^2\bigg|_{S=0}\,,
\end{align}
where we followed the same steps as in the photon exchange example.

\vskip 4pt
As before, we have found the desired correlator up to contact terms. We complete the singular piece at $S = 0$ to the second Legendre polynomial, which is again compatible with the expectation of the flat-space limit
\beq
	\mathop{\rm{lim}}_{{\cal E}=0}A_{4,g} = \mathop{\rm{lim}}_{{\cal E}=0}\frac{1}{{\cal E}^2} \frac{S}{6} P_2 \bigg(\frac{U-T}{S} \bigg) \mapsto \frac{2}{E^3} \bigg( \frac{(u-t)^2}{4s} - \frac{s}{12} \bigg) \,.
\eeq
This fixes the correlator to be 
\be \label{equ:Ag}
\boxed{A_{4,g}=\frac{1}{(S+T+U)^2}\hs\left(\frac{(U-T)^2}{S}-\frac{S}{3}\right) }\ .
\ee 
We will show in Section~\ref{ssec:Int} that the $\tau$-integral \eqref{equ:tpsi-int4} of this Grassmannian correlator reproduces the known correlator in momentum space.

\subsection{Integration Contours}
\label{ssec:Int}

In Section~\ref{ssec:Mandelstams}, we showed that the Mandelstams $S,T,U$ have pairs of zeros at $\tau_i$ and $\bar \tau_i$, for $i=s,t,u$, and  their sum $S+T+U$ vanishes at $\tau=0$. These zeroes lead
to poles in the integrand~$A_4(\tau)$.
To define the integrals in~\eqref{equ:tpsi-int4}, we need to specify integration contours which pick up some of these poles.

\vskip 4pt
We begin by specifying the integration contour for the scalar correlators considered in this section. For these correlators, the function~$A_4$
 has singularities only at  $S=0$, $T=0$, $U=0$, and $S+T+U=0$. We define a contour that encloses the poles at 
$\tau=0,\hs\bar\tau_s,\hs\bar\tau_t,\hs\bar\tau_u$ in the counter-clockwise direction. Evaluating the $\tau$-integral along this contour therefore amounts to summing the residues at these poles.

\begin{figure}[t!]
	\centering
	\includegraphics[scale=1]{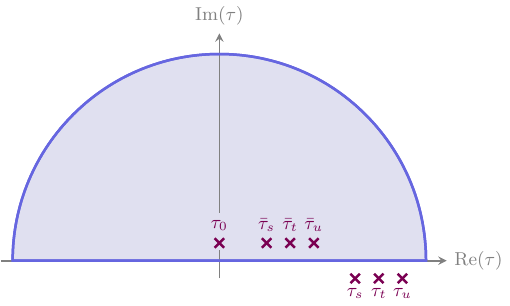} 
	\caption{Illustration of the possible poles in the $\tau$-plane and the choice of contour used in this section.}
	\label{fig:contour-tau}
\end{figure}

\vskip 4pt
Interestingly, this contour can be interpreted by assigning the following  $i\epsilon$-prescription for the Mandelstams:
\be \label{equ:ieps-mandelstams}
\begin{aligned}
	S &\mapsto S+i\epsilon\,,\\
	T &\mapsto T+i\epsilon\,,\\
	U &\mapsto U+i\epsilon\,,
\end{aligned}
\ee 
with $\epsilon>0$. This positive imaginary part of the Mandelstams implies imaginary parts for the roots \eqref{equ:tau-s} and \eqref{equ:tau-sb} given by\footnote{Here, we implicitly assume a $(2,2)$ signature, in which all spinor helicity variables and minors are real, as well as~$E,\hs k_s,\hs k_t,\hs k_u>0$. Consequently, the roots $\tau_i$ and $\bar \tau_{i}$ are also real.}
\be  
\begin{aligned}
	\tau_s &\mapsto \tau_s-i\epsilon\,, \qquad \tau_t \mapsto \tau_t-i\epsilon\,, \qquad \tau_u \mapsto \tau_u-i\epsilon\,,  \\
	\bar \tau_{s} &\mapsto \bar \tau_{s}+i\epsilon\,, \qquad \bar \tau_{t} \mapsto \bar \tau_{t}+i\epsilon\,, \qquad \bar \tau_{u} \mapsto \bar \tau_{u}+i\epsilon\,.
\end{aligned}
\ee 
Furthermore, it gives a positive imaginary part to the root $\tau_0= i\epsilon$ of $(S+T+U)(\tau)$. Hence, if the Grassmannian correlator has poles only at $S=0$, $T=0$, $U=0$, and  $S+T+U=0$, then its singularity structure is that shown in Figure~\ref{fig:contour-tau}. Moreover, as shown in this figure, the contour that encloses the poles at $\bar \tau_{s}+i\epsilon$, $\bar \tau_{t}+i\epsilon$, $\bar \tau_{u}+i\epsilon$ and $\tau_0=+i\epsilon$ is simply given by the real line together with the semi-arc in the upper half-plane.\footnote{Notice that the integral along the semi-arc may not vanish if the residue at $\tau=\infty$ is non-trivial, and thus it may not be ignored in our contour.} For correlators that satisfy the homogeneous Ward identity (such as the correlators associated to photon and graviton exchange), we only integrate \eqref{equ:main-integral} over the right branch of the Grassmannian. For correlators that obey the inhomogeneous Ward identity (such as the correlators of Yang--Mills theory), on the other hand, the integral is performed over both branches.

\subsubsection*{Photon exchange}

In Section~\ref{ssec:factorization}, we found the following
Grassmannian correlator for the case of four conformally coupled scalar exchanging a photon in the $s$-channel:
\be \label{equ:A4-gamma-2}
A_{4,\gamma} = \frac{1}{S+T+U}
\hs\frac{T-U}{S}\,.
\ee 
To derive the corresponding momentum-space correlator, we
perform the integral \eqref{equ:tpsi-int4} using the contour shown in Figure~\ref{fig:contour-tau}. In this case, the integral amounts to adding up the residues of $A_{4,\gamma}(\tau)$ at $\tau=0$ and $\tau=\bar \tau_s$, which are straightforward to compute. The result is
\beq
\psi_{4,\gamma}=\int \frac{\ud\tau}{2\pi i} \hs A_{4,\gamma} = \frac{1}{EE_LE_R} \left( \frac{k_{12}\hs k_{34}\hs\alpha\hs\beta+k_s^2\hs(k_u^2-k_t^2)}{2k_s^2} \right) -  \frac{1}{E} \hs \frac{\alpha\hs\beta}{2k_s^2} \,,
\label{equ:psi4-gamma-2}
\eeq
where 
\be  
\begin{aligned}
k_{12} \equiv k_1+k_2\,, \qquad \alpha \equiv k_1 - k_2\,, \qquad k_s &\equiv |\vec k_1+\vec k_2|\,,  \\
k_{34} \equiv k_3+k_4\,,\qquad \beta \equiv k_3 - k_4\,, \qquad k_t &\equiv |\vec k_1+\vec k_4|\,, \\
k_u &\equiv |\vec k_1+\vec k_3|\,.
\end{aligned}
\ee 
This is precisely the known correlator of four conformally coupled scalars exchanging a photon in the $s$-channel; cf.~equation~(4.55) in \cite{Arkani-Hamed:2018kmz}. It is remarkable how simple this object looks at the level of the cosmological Grassmannian~\eqref{equ:A4-gamma-2}. Indeed, it is practically as simple as the corresponding flat-space amplitude.

\subsubsection*{Graviton exchange} 

Next, we consider the case of four scalars exchanging a graviton in the $s$-channel. 
In Section~\ref{ssec:factorization}, we found the following Grassmannian correlator 
\be \label{equ:A4-g}
\begin{aligned}
A_{4, g} &= \frac{1}{(S+T+U)^2}
\left(\frac{(U-T)^2}{S}- \frac{S}{3}\right)  .
\end{aligned}
\ee
As before, we perform the integral \eqref{equ:tpsi-int4} using the contour shown in Figure~\ref{fig:contour-tau}, which encloses the poles at $\tau=0$ and $\tau=\bar \tau_s$. Adding up their residues yields
\beq\label{equ:psi4-g}
\psi_{4,g} =-  \frac{(k_s E +E_LE_R)k_s^4 }{3\hs E^3E_L^2E_R^2} \Pi_{2,2} +  \frac{k_s^2}{3\hs E^3} \Pi_{2,1} -\frac{(E_LE_R-k_s E)}{3\hs E^3} \Pi_{2,0}\,,
\eeq
where the polarization sums are
\be
\begin{split}
\Pi_{2,2}&\equiv\frac{3\hs \left[2\hs (k_s^2 \hs (k_u^2-k_t^2)+k_{12}\hs k_{34}\hs \alpha\hs \beta)^2-(k_s^2-k_{12}^2)\hs(k_s^2-k_{34}^2)\hs(k_s^2-\alpha^2)\hs(k_s^2-\beta)\right]}{4\hs  k_s^8}\,,\\
\Pi_{2,1}&\equiv\frac{3\hs \alpha\hs \beta\hs (k_s^2 \hs(k_u^2-k_t^2)+k_{12}\hs k_{34}\hs \alpha\hs \beta)}{k_s^6}\,,\\
\Pi_{2,0}&\equiv\frac{(k_s^2-3\hs \alpha^2)\hs(k_s^2-3\hs\beta^2)}{4\hs k_s^4}\,.
\end{split}
\ee 
The result~\eqref{equ:psi4-g} is exactly the known momentum-space correlator; cf.~equation (4.34) in~\cite{Baumann:2021fxj}. The simplicity of the Grassmannian formula~\eqref{equ:A4-g} compared to its momentum-space counterpart~\eqref{equ:psi4-g} is even more striking than the photon exchange case. As before, it is virtually as simple as the corresponding scattering amplitude.

\section{Yang--Mills Correlator}
\label{sec:YM} 
A major breakthrough in our understanding of scattering amplitudes was the discovery by Parke and Taylor of a remarkably compact formula for the scattering of $n$ gluons in the maximally helicity-violating configuration~\cite{Parke:1986gb}:
\beq
M(1^- 2^- 3^+ \cdots n^+) = \frac{\langle 12\rangle^4}{\langle 12\rangle \langle 23\rangle \cdots \langle (n-1) n \rangle \langle n1\rangle}\,.
\label{equ:ParkeTaylor}
\eeq
We are still far from an analogous formula in the cosmological context. As a modest step in this direction, we will now use the Grassmannian formalism to derive a simple expression for the four-point function of pure Yang--Mills theory. We first show how this correlator can be bootstrapped directly in Grassmannian space using the principles outlined in Section~\ref{sec:4pt}, and then verify, by explicit integration, that it reproduces the correct momentum-space result. We will see that the Grassmannian representation is considerably simpler than its momentum-space counterpart.

\subsection{Grassmannian Space}

We will focus on the $--++$ color-ordered correlator. Unlike in the previous examples, the channel-by-channel decomposition of the four-point function is not physical, and hence we must compute the sum that arises from the $s$ and $t$-channel poles. Using the factorization property~\eqref{equ:Res-S=0} and the results of Section~\ref{sec:gaugeinv3pt}, the residue of the $s$-channel pole is
\begin{align}
	\mathop{\rm{Res}}_{S=0}A(1^- 2^- 3^+ 4^+) &=\frac{(1  2\bar I_s)^2}{(2 I_s\bar I_s)(I_s \bar 2\bar I_s)}\hs \frac{(I_s \bar 3\bar 4)^2}{(3 \bar 4 4)(4 \bar 3\bar 4)} + \frac{(1 2 I_s)^2}{(2 \bar I_s I_s)(I_s \bar 2\bar I_s)}\hs \frac{(\bar I_s \bar 3\bar 4)^2}{(3 4\bar 4)(4 \bar 3\bar 4)} \nonumber\\[4pt]
	&=-\frac{(1 2\bar 3 \bar4)^2}{(2 \bar3 \bar 4 4)(\bar2 3\bar 44)}\bigg|_{S=0}=\frac{4\hs ( 1 2\bar 3 \bar4)^2}{(S+T+U)(T-U)}\bigg|_{S=0}\,.
	\label{equ:sRes}
\end{align}
Going to the second line, we have added $2(12\bar I_s)(12 I_s)(I_s\bar3\bar4)(\bar I_s\bar3\bar4) = 0$ to complete the square.  A similar computation for the residue of the $t$-channel pole yields
\beq
	\mathop{\rm{Res}}_{T=0}A(1^- 2^- 3^+ 4^+) = \frac{4\hs (12\bar 3 \bar4)^2}{(S+T+U)(S-U)}\bigg|_{T=0}\,.
	\label{equ:tRes}
\eeq
We note that the ``gluing" of three-point functions in the factorization limit has led to novel singularities of the Grassmannian correlator at $T=U$ (on the $s$-channel pole) and at $S=U$ (on the $t$-channel pole). 
To construct the correlator away from the factorization limits, we impose that it has only simple poles at $S=0$, $T=0$ and $S\pm T\pm U = 0$. A color-ordered correlator must further satisfy  the {\it Kleiss--Kuijf relation}~\cite{Kleiss:1988ne}
\be 
A(1^- 2^- 3^+ 4^+) +A(1^-  3^+ 4^+2^-) +A(1^- 4^+ 2^- 3^+) =0\,,
\label{equ:KK}
\ee
which is simply a consequence of the Jacobi identity of the color structures. The following function is consistent with the above factorization properties, permutation symmetry in $S\leftrightarrow T$ and the Kleiss--Kuijf relation \eqref{equ:KK}:
\be \label{equ:AYM-kk}
A(1^- 2^- 3^+ 4^+) = \frac{(12\bar 3\bar 4)^2}{ST}\left(\frac{3}{S+T+U}+\frac{1}{S+T-U}-\frac{1}{-S+T+U}-\frac{1}{S-T+U}\right) .
\ee 
Remarkably, the first term takes exactly the same form as the corresponding amplitude (written in terms of the Mandelstams $S,T,U$), while the other terms arise from sign flips of $S$, $T$ and $U$. The flat-space limit \eqref{equ:sub} leads to 
\be 
\lim_{\mathcal{E}\to0} A(1^- 2^- 3^+ 4^+)\propto \lim_{\mathcal{E}\to0} -\frac{1}{\mathcal{E}}\hs\frac{(1 2 \bar 3 \bar 4)^2}{S T}   \mapsto \frac{1}{E} \bigg(\hspace{-3pt}- \frac{\langle 12\rangle^2[34]^2}{s\hs t}\bigg) \,,
\ee 
which contains the correct Yang--Mills amplitude.

\vskip 4pt
In the following, we will also find it convenient to define the {\it reduced} color-ordered correlator as the part of  \eqref{equ:AYM-kk} that is symmetric under the exchange $U\mapsto -U$. Writing $\hat A \equiv \frac{1}{2}\big[A+ A({U\mapsto-U})\big]$, we get
\beq
\label{eq:posterchild}
\boxed{\hat A(1^- 2^- 3^+ 4^+) = 2\, \frac{(1  2 \bar 3 \bar 4)^2}{S T} \left(\frac{1}{S+T+U} + \frac{1}{S+T-U}\right) } \ .
\eeq
This form of the Grassmannian correlator is convenient because its integral leads to a simple discontinuity $\mathop{\rm{Disc}}_{k_1^2,k_3^2}[\psi]$ in momentum space.\footnote{The fact that we obtain this specific discontinuity is related to the choice of the Grassmannian correlator~\eqref{eq:posterchild} and its underlying exchange symmetry $U \mapsto -U$. Other choices for the Grassmannian correlator would lead to other discontinuities. 
For instance, the correlator \eqref{equ:AYM-kk} that complies with the Kleiss--Kuijf relation can be integrated to give the triple discontinuity $\mathop{\rm{Disc}}_{k_1^2,k_2^2,k_3^2}[\psi]$ (see Section~\ref{sec:GR-momentum}). 
It would be interesting to understand systematically how these different discontinuities are represented in the Grassmannian.} As we will show in the next section, this discontinuity  can be integrated
to produce the desired correlator, and thus it contains all its information.\footnote{Although not shown explicitly in this paper, we have verified that all correlators solving the \textit{inhomogeneous} Ward identity (such as $\langle JOJO \rangle$, $\langle TOTO \rangle$, $\langle TTTT \rangle$) 
exhibit novel singularities besides the usual poles at $S = T=U=0$ and $S+T+U=0$. We believe that these singularities are again related to the fact that we are solving the \textit{homogeneous} Ward--Takahashi identity~\eqref{equ:sct-hom}, and therefore only produce discontinuities of correlators.}

\subsection{Momentum Space}
\label{ssec:dispersive}

The Yang--Mills correlator in momentum space is~\cite{Albayrak:2018tam, Baumann:2020dch} 
\beq
\psi(1^{h_1} 2^{h_2} 3^{h_3} 4^{h_4}) = \mathcal{T}_1 \mathcal{F}_1 + \mathcal{T}_2 \mathcal{F}_2 + \mathcal{T}_3 \mathcal{F}_3 + (\text{$t$-channel})\,,
\eeq
where the ``tensor structures'' $\mathcal{T}_a$ are
\beq
\begin{aligned}
	\mathcal{T}_1 &\equiv 4(\vec \epsilon_1 \cdot \vec \epsilon_3) (\vec \epsilon_2 \cdot \vec \epsilon_4)- 2(\vec \epsilon_1 \cdot \vec \epsilon_2)(\vec \epsilon_3 \cdot \vec \epsilon_4) - 2(\vec \epsilon_1 \cdot \vec \epsilon_4)(\vec \epsilon_2 \cdot \vec \epsilon_3) \,,\\[4 pt]
	\mathcal{T}_2 & \equiv 8\Big[(\vec \epsilon_3 \cdot \vec k_1)(\vec \epsilon_4 \cdot \vec k_2) - (\vec \epsilon_4 \cdot \vec k_1)(\vec \epsilon_3 \cdot \vec k_2) \Big] (\vec \epsilon_1 \cdot \vec \epsilon_2)  \\
	& + 8 \Big[(\vec \epsilon_1 \cdot \vec k_3)(\vec \epsilon_2 \cdot \vec k_4)-(\vec \epsilon_2 \cdot \vec k_3)(\vec \epsilon_1 \cdot \vec k_4)\Big](\vec \epsilon_3 \cdot \vec \epsilon_4) \\
	& + 8 \Big[ (\vec \epsilon_1 \cdot \vec k_2) \, \vec \epsilon_2 - (\vec \epsilon_2 \cdot \vec k_1) \, \vec \epsilon_1 \Big] \cdot \Big[ (\vec \epsilon_3 \cdot \vec k_4) \, \vec \epsilon_4 - (\vec \epsilon_4 \cdot \vec k_3) \, \vec \epsilon_3 \Big] \,, \\[4pt]
	\mathcal{T}_3 &\equiv 2(\vec \epsilon_1 \cdot \vec \epsilon_2)(\vec \epsilon_3 \cdot \vec \epsilon_4) \,,
\end{aligned}
\eeq
and the ``form factors'' $\mathcal{F}_a$ are
\beq
\begin{aligned}
	\label{eq:formfactors}
	\mathcal{F}_1 &\equiv \frac{1}{E}\,, \quad
	\mathcal{F}_2  \equiv \frac{1}{E E_L E_R}\,, \\[4pt]
	\mathcal{F}_3 &\equiv \frac{1}{EE_LE_R} \left( \frac{k_{12}\hs k_{34}\hs\alpha\hs\beta+k_s^2\hs(k_u^2-k_t^2)}{k_s^2} \right) - \frac{1}{E} \hs \frac{\alpha\hs\beta}{k_s^2} .
\end{aligned}
\eeq
We will now show that we recover this result by performing the integral \eqref{equ:tpsi-int4} of the Grassmannian correlator~\eqref{eq:posterchild}. We again consider the contour shown in Figure~\ref{fig:contour-tau}. Depending on the choice of external kinematics, this amounts to adding the residues at $\tau = 0, \tau = \bar \tau_s, \tau = \bar \tau_t$ and either of the two poles of $S+T-U$. We denote the latter as  
\beq 
\tau_{13} \equiv \frac{E_L^{(u)} \bar E_L^{(u)}}{\langle \bar1\bar3\rangle \langle \bar2\bar4 \rangle E} \quad  {\rm and} \quad \tau_{24} \equiv \frac{E_R^{(u)} \bar E_R^{(u)}}{\langle \bar1\bar3\rangle \langle \bar2\bar4 \rangle E} \, ,
\eeq
where $E_{L,R}^{(u)}$ are the left and right energies in the $u$-channel, while $\bar E_{L,R}^{(u)}$ are obtained from the former by flipping the sign of $k_u$.

\vskip 4pt
We first consider the case $k_{24} - k_{13} > 0$, which selects the pole at $\tau = \tau_{13}$. Using \eqref{eq:posterchild}, we then obtain  
\beq\label{equ:disck1k3}
\int \frac{\ud\tau}{2\pi i} \: \hat A(1^- 2^- 3^+ 4^+) = \mathop{\rm{Disc}}_{k_1^2,k_3^2}\hs\big[\psi(1^- 2^-3^+ 4^+)\big]\,.
\eeq
Given the general decomposition of a momentum-space correlator into a sum of products of tensor structures and form factors, the discontinuity with respect to a given energy $k_m^2$ \eqref{equ:Disc-def} is more concretely defined as 
\beq
	\mathop{\rm{Disc}}_{k_m^2} \big[ \psi \big] = \sum_a \mathcal{T}_a(\vec \epsilon_i \cdot \vec \epsilon_j, \vec \epsilon_i \cdot \vec k_j) \Big[\mathcal{F}_a (k_{i \neq m}, k_m) - \mathcal{F}_a (k_{i \neq m}, -k_m)\Big]\,.
	\label{eq:disc-def2}
\eeq
Equivalently, one may view the discontinuity as acting on each form factor separately. 
Similarly, when $k_{24} - k_{13} < 0$,  the contour instead selects the pole at $\tau=\tau_{24}$, and the integral yields $\mathop{\rm{Disc}}_{k_2^2,k_4^2}\hs\big[\psi(1^- 2^- 3^+ 4^+)\big]$.  We see that the result is a discontinuity  obtained by a flip of the energies of either the 
$\{1,3\}$ or $\{2,4\}$ pair of particles, which is ultimately related to the $U \mapsto -U$ exchange symmetry in Grassmannian space.\footnote{In particular, note that the discriminant of $S+T-U$ as a quadratic polynomial in $\tau$ is $4(k_{24}-k_{13})^2$, while the discriminant of $S+T+U$ is instead $4(k_{24}+k_{13})^2$. Hence, they are related by a flip of the energies $k_1,k_3$ in momentum space, or by a flip of the sign of $U$ in Grassmannian space.} This provides an a posteriori justification for our choice of Grassmannian correlator.

\vskip 4pt
Although we have only produced a discontinuity of the momentum-space correlator, we will now demonstrate that it nevertheless contains all information about the full correlator. To substantiate this claim, we show that the discontinuities discussed above can be straightforwardly inverted without any additional input---unlike other types of cuts, such as those involving internal legs, which generally require supplementary information~\cite{Ansari:2025fvi}.\footnote{The results of~\cite{Ansari:2025fvi}, formulated in terms of Schwinger parameters, are supported only on the $S=0$ locus of our cosmological Grassmannian, whereas our results extend beyond this locus. Indeed, as shown in Appendix~\ref{app:factorization}, taking the discontinuity with respect to the internal energy $k_s$ localizes the Grassmannian integral to $S=0$.} 
The idea is to perform a dispersive integral on $\mathop{\rm{Disc}}_{k_1^2,k_3^2} \big[ \mathcal{F}_a \big]$ that returns $\mathcal{F}_a$ for all the form factors, as was done in \cite{Baumann:2024ttn} for $n = 3$ spinning correlators. Up to trivial permutations in the $t$-channel, one must recover the three form factors in~\eqref{eq:formfactors}. To reproduce this, we consider the following \textit{dispersive integral}
\beq
\label{eq:dispint}
\tilde {\mathcal{F}}_a(k_i) = \frac{1}{(2\pi i)^2} \int_{-\infty}^{+\infty} \bigg(\prod_{i=1,3}  \frac{\omega_i \, \ud\omega_i}{\omega_i^2-(k_i-i\varepsilon)^2} \bigg)\, \mathop{\rm{Disc}}_{k_1^2,k_3^2} \big[ \mathcal{F}_a \big](\omega_1, k_2-i\varepsilon, \omega_3, k_4-i\varepsilon) \,,
\eeq
where $\varepsilon >0$ is a small regulator. For $a = 1,2,$ the integral can be straightforwardly computed using the residue theorem by closing the contour in the lower half-plane for both $\omega_1$ and $\omega_3$, which shows that $\tilde{\mathcal{F}}_{1,2} = \mathcal{F}_{1,2}$, as desired. The third form factor, on the other hand, requires a careful subtraction procedure. We first split $\mathcal{F}_3$ into three sub-factors:
\beq
\begin{aligned}
	\mathcal{F}_3^{(1)} & = \frac{1}{E E_L E_R} \left( \frac{k_2^2 \beta + k_4^2 \alpha}{k_s} -\alpha \beta - k_t^2 + k_u^2\right) , \\[4pt]
	\mathcal{F}_3^{(2)} & = -\frac{1}{E E_L E_R} \frac{k_1^2 \beta}{k_s} \,, \quad \mathcal{F}_3^{(3)} = -\frac{1}{E E_L E_R} \frac{k_3^2 \alpha}{k_s} \,.
\end{aligned}
\eeq
 This split is demanded by the different behavior of the sub-factors as $k_1, k_3 \to \infty$. For the dispersive integral \eqref{eq:dispint} to correctly recover the form factor, it is necessary for $\mathcal{F}_a$ as a function of $\omega_i^2$ to have a vanishing residue at $\omega_{1,3}^2 = \infty, $\footnote{We refer to Appendix C of \cite{Baumann:2024ttn} for details.} which is the case for $\mathcal{F}^{(1)}_3$, but not for $\mathcal{F}^{(2,3)}_3$. Luckily, this can be solved by a simple subtraction that softens the behavior at infinity, i.e. it holds that:
\beq
\begin{aligned}
	\label{eq:dispintnew}
	{\mathcal{F}}^{(2)}_3(k_i) & = \frac{1}{(2\pi i)^2} \int_{-\infty}^{+\infty} \bigg(\prod_{i=1,3}  \frac{\omega_i \, \ud\omega_i}{\omega_i^2-(k_i-i\varepsilon)^2} \bigg) \frac{k_1^2}{\omega_1^2} \, \mathop{\rm{Disc}}_{k_1^2,k_3^2} \big[ \mathcal{F}_3^{(2)} \big](\omega_1, k_2-i\varepsilon, \omega_3, k_4-i\varepsilon) \,, \\[4pt]
	{\mathcal{F}}^{(3)}_3(k_i) & = \frac{1}{(2\pi i)^2} \int_{-\infty}^{+\infty} \bigg(\prod_{i=1,3}  \frac{\omega_i \, \ud\omega_i}{\omega_i^2-(k_i-i\varepsilon)^2} \bigg) \frac{k_3^2}{\omega_3^2} \, \mathop{\rm{Disc}}_{k_1^2,k_3^2} \big[ \mathcal{F}_3^{(3)} \big](\omega_1, k_2-i\varepsilon, \omega_3, k_4-i\varepsilon) \,.
\end{aligned}
\eeq
We stress that, even though a subtraction procedure was necessary, it did not introduce a pole at $\omega_{1,3} = 0$ thanks to the soft behavior of the sub-factors, so no extra physical information was needed. Summing everything back together yields the desired Yang--Mills correlator. 

\section{Conclusions}
\label{sec:conclusions}

It has long been a challenge in cosmology that correlators of massless spinning fields appear surprisingly complicated. 
In this paper, we have, for the first time, found a representation that reveals the true simplicity of these correlators.

\vskip 4pt
We uncovered the hidden simplicity of spinning (A)dS correlators by writing them as integrals over the orthogonal Grassmannian:
\beq
\psi_n(\Lambda) = \int \ud C\, \delta(C\cdot \Lambda) \,A_n(C)\,,
\label{equ:Grass-Integral}
\eeq
where $\Lambda$ is a $2n\times 2$ matrix of spinor helicity variables encoding the external momenta and $C$ is a $n \times 2 n$ matrix made from the Grassmannian coordinates. The three-point functions $A_3(C)$ are fixed by little group covariance alone, while the four-point functions $A_4(C)$ are determined by their factorization properties. Exploring a number of non-trivial examples, we found remarkably simple expressions for the Grassmannian correlators.

\vskip 4pt
This work marks an initial step toward a comprehensive framework for cosmological correlators in Grassmannian space. Several natural avenues for future research suggest themselves:
\begin{itemize}
\item The integral \eqref{equ:Grass-Integral} solves the {\it homogeneous} Ward--Takahashi identity and therefore captures only a specific discontinuity of the full correlator. Nevertheless, with a suitable choice of contour, it encodes all the information of the full correlator, which can be reconstructed through a dispersive integral.  It would be desirable, however, to find a deformation of our Grassmannian formalism that eliminates this two-step procedure and instead produces the full correlator directly. 
\item  In Appendix~\ref{app:gravity}, we present a preliminary investigation of of the graviton four-point function in Grassmannian space. As for the Yang--Mills correlator, its functional form is drastically simpler than its momentum-space counterpart. 
We show that the Grassmannian answer leads to a triple discontinuity of the wavefunction coefficient in momentum space, but don't perform the additional dispersive integral to reproduce the full correlator. We hope to return to a more systematic investigation of the Grassmannian formulation of graviton correlators in future work. 


\item The examples in this paper were limited to $n$-point functions with $n=3$ and $4$. An important next step is to extend our treatment to higher-point functions, potentially employing recursion relations to construct them from the known lower-point results.

\item 
It would be nice to clarify the relation between the cosmological Grassmannian and Mellin space, which has also been used to express conformal correlators in a manner closely analogous to scattering amplitudes~\cite{Penedones:2010ue,Paulos:2011ie}. 
It is natural to ask whether the Mellin-space Mandelstam variables are related to the Mandelstams introduced in this work.

\item It would be interesting to develop a supersymmetric extension of our Grassmannian formalism, with a focus on ABJM correlators. Since the factorization properties hold for tree-level interactions in  AdS$_4$, the formalism may provide access to strongly coupled correlators. By contrast, the same orthogonal Grassmannian OGr$(n,2n)$, with a different bilinear form $Q$, describes ABJM amplitudes at weak coupling~\cite{Huang:2013owa}. It is therefore intriguing to ask whether these two descriptions can be interpolated within the Grassmannian framework.

\item Finally, we would like to make closer contact with the ideas of positive geometry. This may require restricting to theories with bulk conformal symmetry, such as Yang–Mills theory, where a more constrained Grassmannian could make the notion of positivity essential.\footnote{We thank Jaroslav Trnka for emphasizing this point to us.} If such a geometry exists, several natural questions arise: What is the precise interpretation of the positive regions of the orthogonal Grassmannian? Can integrals over the Grassmannian be reformulated as integrals of canonical forms associated with this geometry, embedding the construction within the broader framework of positive geometries?
\end{itemize}
It is remarkable that the Grassmannian emerges so naturally as the framework in which all kinematic constraints of spinning (A)dS correlators can be made manifest. We believe this marks only the beginning of a deeper exploration of the geometric origins of these physical objects, and we look forward to seeing where these investigations will lead.

 \vspace{0.2cm}
 \paragraph{Acknowledgments} We thank Tim Adamo, Nima Arkani-Hamed, Veronica Calvo Cortes, Mariana Carrillo-González, Lorenzo di Pietro, Harry Goodhew, Subramanya Hegde, Yu-tin Huang, Austin Joyce, Théo Keseman, Joris Koefler, Hayden Lee, Nat Levine, Lionel Mason, Ugo Moschella, Julio Parra-Martinez, Miguel Paulos, Balt van Rees, Slava Rychkov, David Skinner, Charlotte Sleight, Andy Strominger, Bernd Sturmfels, Massimo Taronna, Jaroslav Trnka and Alessandro Vichi for insightful discussions. 
  
 \vskip 4pt
 MA presented preliminary results of this work at the ``UNIVERSE+ Annual Collaboration Meeting" at Ringberg Castle and the workshop ``Cosmology Meets Non-Linear Algebra" in Amsterdam, while GLP spoke at the ``Celestial Meets de Sitter Holography" meeting in Capri, the joint Belgian hep-th seminar, and the INFN-FLAG national meeting. They thank the participants of these meetings for their valuable feedback.
 
\vskip 4pt
The research of DB and MA is funded by the European Union (ERC,  \raisebox{-2pt}{\includegraphics[height=0.9\baselineskip]{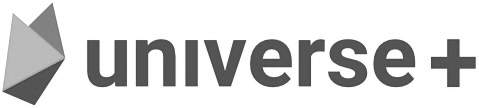}}, 101118787). DB is further supported by a Yushan Professorship at National Taiwan University (NTU) funded by the Ministry of Education (MOE) NTU-112V2004-1. He also holds the Chee-Chun Leung Chair of Cosmology at NTU. GLP and FR are supported by the ERC (NOTIMEFORCOSMO, 101126304), by Scuola Normale, and by INFN (IS GSS-Pi). The research of GLP is moreover supported by the Italian Ministry of Universities and Research (MUR) under contract 20223ANFHR (PRIN2022). MHGL is supported by a postdoctoral fellowship at NTU funded by the Ministry of Education (114V2004-3).

\appendix
\newpage
\section{A Brief Introduction to Twistors}
\label{app:twistors}

In Section~\ref{ssec:twistors}, we explained how our Grassmannian correlators arise from twistor space. As background for that discussion, this appendix provides a brief introduction to twistors and their recent application to cosmological correlators~\cite{Baumann:2024ttn}.

\subsection{Twistors for Pedestrians}
 
Geometrically, a twistor can be thought as a null plane in the complexified bulk spacetime (A)dS$_4$, that reduces to a null line in its three-dimensional boundary. Thus, in some sense twistors lie between position and momentum space: They are not localized at spacetime points or at fixed momenta, but instead on null planes.

\vskip 4pt
Twistors can be written as a complex four-component spinor defined up to scale, $Z^A \sim \rho Z^A$, with $\rho \in \mathbb{C}$. 
In terms of two-component Weyl spinors, we have
\begin{equation}
	Z^A = (\lambda^\alpha,\mu_{\dot\alpha})\,, 
\end{equation}
where $A = 1,2,3,4$ and $\alpha, \dot\alpha = 1,2$.
Given a \textit{complexified} boundary position $x^\mu \in \mathbb{C}^3$ at $\eta = 0$ (dS$_4$) or $z = 0$ (AdS$_4$),\footnote{Talking about AdS$_4$ or dS$_4$ in different signatures effectively amounts to different slices of a common complexified spacetime.} we can associate a $2 \times 2$ matrix $x_{\alpha \dot\alpha} = x_\mu \sigma^\mu_{\alpha \dot\alpha}$ to it.
For this spinor~$Z^A$ to correspond a twistor, we introduce the following \textit{incidence relation}:
\be \label{equ:incidence}
\mu_{\dot\alpha} = x_{\alpha\dot\alpha} \lambda^\alpha\,.
\ee 
This is a nonlocal relation between boundary positions and twistor space that maps a fixed spinor~$Z^A$ to all the positions such that \eqref{equ:incidence}
holds, which form a null line in the complexified position space $\mathbb{C}^3$. Conversely, the incidence relation maps each position to a (projective) $2$-plane in twistor space, parametrized by $\lambda^\alpha$ up to a scale. 

\vskip 4pt
To specialize to the case of interest, we can take a slice of  the complexified spacetime: this is usually taken to be the Euclidean slice (no timelike coordinates), the Lorentzian slice (one timelike coordinate) or the ``split signature'' slice (two timelike coordinates). It is convenient to study the latter, and then analytically continue our results to the others. In terms of our preferred dS$_4$ geometry, this means that one of the spatial directions is also taken to be a time direction along with $\eta$, so that the future boundary $\mathcal{I}^+$ is $\mathbb{R}^{1,2}$ rather than $\mathbb{R}^3$. We refer to the resulting spacetime as dS$_{2,2}$.

\vskip 4pt
Another essential ingredient are the \textit{dual twistors} $W_A = (\bar \mu_\alpha, \bar \lambda^{\dot \alpha})$, which can be interpreted as the conjugate variables of the twistors $Z^A$, just like momenta are dual to positions. Explicitly, any function of a twistor $G(Z^A)$ can be converted to dual twistor space by a Fourier transform:
\begin{equation}
	\tilde G(W_A) = \int \ud^4 Z \, \exp(i Z^A W_{A}) \, G(Z^A)\,,
\end{equation}
where $Z^A W_A = \lambda^\alpha \bar \mu_\alpha + \bar \lambda^{\dot \alpha} \mu_{\dot\alpha}$.

\subsection{Twistors for Cosmology}

The utility of (dual) twistors is twofold. First, twistors realize the spacetime isometries in a natural way. In particular, by associating a twistor $Z_i^A$ to each particle $i$, any function is automatically $\mathrm{SO}(2,3)$-invariant if it only depends on inner products of these twistors:
\begin{equation}\label{equ:twistorcontractions}
	\begin{gathered}
		F(Z_i^A) = F(Z_i^A \Omega_{AB} Z_j^B) \equiv F(Z_i \cdot Z_j)\,, \quad \Omega_{AB} = \begin{pmatrix}
			0 & 1_{2\times 2} \\
			- 1_{2 \times 2} & 0
		\end{pmatrix} .
	\end{gathered}
\end{equation}
Here, $\Omega_{AB}$ is a symplectic form that is often referred to as the ``infinity twistor'' because it encodes information on the structure of spacetime at infinity.
Note that, unlike for flat space, $\Omega_{AB}$ allows the contraction between dotted and undotted indices using $\delta^{\dot\alpha}_\alpha$. In the following, all indices will always be undotted.  We can associate either a twistor or a dual twistor to a particle, so that we can also have the contractions $Z_i \cdot W_{j}$ and $W_{i} \cdot W_{j} = W_{i,A} \Omega^{AB} W_{j,B}$, with $\Omega^{AB} = -\Omega_{AB}$.

\vskip 4pt
The second important feature of (dual) twistors is that they naturally describe 
boundary currents in a way that trivializes conservation.
Concretely, a spin-$\ell$ current can be written as the following \textit{Penrose transform}:
\begin{equation}
	\begin{aligned}
		J^{\alpha_1 \cdots \alpha_{2\ell}}(x^\mu) & = \int \langle \lambda \, \dif \lambda \rangle \: \lambda^{\alpha_1} \cdots \lambda^{\alpha_{2\ell}} \, J(Z^A) \bigg|_{\mu_{\alpha} = x_{\alpha \beta} \lambda^\beta}\,, \\
		\tilde J^{\alpha_1 \cdots \alpha_{2\ell}}(x^\mu) & = \int \langle \bar \lambda \, \dif \bar \lambda \rangle \: \frac{\partial}{\partial \bar \mu_{\alpha_1}} \cdots \frac{\partial}{\partial \bar \mu_{\alpha_{2\ell}}} \, \tilde J(W_A) \bigg|_{\bar \mu_{\alpha} = x_{\alpha \beta} \bar \lambda^\beta}\,,
	\end{aligned}
\end{equation}
where $\langle \lambda \, \dif \lambda \rangle \equiv \lambda^\alpha \varepsilon_{\alpha\beta} \dif \lambda^\beta$ and $\langle \bar\lambda \, \dif \bar\lambda \rangle \equiv \bar\lambda^\alpha \varepsilon_{\alpha\beta} \dif \bar\lambda^\beta$. The above integrals are only projectively well-defined if 
\beq
\begin{aligned}
J(\rho Z^A) &= \rho^{-2\ell-2} J(Z^A)\,, \\
\tilde J(\rho W_A) &= \rho^{2\ell-2} \tilde J(W_A)\,,
\end{aligned}
\eeq
which implies that the scaling dimension of the current is $\Delta = \ell+1$ (see Section~\ref{sec:current}).
This Penrose transform allows us to write correlators of $n$ conserved currents in position space. 
For example, assigning a twistor to each current, we can write 
\begin{equation}
	\langle J_1 (x_1^\mu) \cdots J_n (x_n^\mu) \rangle = \int \prod_{i=1}^n \langle \lambda_i \, \dif \lambda_i \rangle (\zeta_i\cdot \lambda_i)^{2\hs\ell_i}\, F_n(Z_i \cdot Z_j) \bigg|_{\mu_{i,\alpha} = x_{i,\alpha\beta} \lambda_i^\beta} \,,
\end{equation}
where $J_i(x_i^\mu)\equiv \zeta_{i,\alpha_1}\cdots \zeta_{i,\alpha_{2\ell_i}} J^{\alpha_1 \cdots \alpha_{2\ell_1}} (x_i^\mu)$, with $\zeta_{i,\alpha}$ being polarization vectors. The resulting correlator is manifestly conserved and invariant under three-dimensional conformal transformations. 

\vskip 4pt
Finally, let us mention that it is convenient to write these correlators in twistor space as
\be 
F_n(Z_i\cdot Z_j)= \int \ud c_{ij} \: A_n(c_{ij}) \: \exp \left( \frac{i}{2} c_{ij}Z_i \cdot Z_j \right) ,
\ee 
where $c_{ij}$ are the so-called Schwinger parameters. The reason is that the form of $A_n(c_{ij})$ is much simpler than $F_n(Z_i\cdot Z_j)$. Indeed, as explained in Section \ref{ssec:twistors}, these $c_{ij}$ parameterize the cosmological Grassmannian and the functions $A_n(c_{ij})$ are simply rational instead of distributional.

\newpage
\section{Factorization from Unitarity} 
\label{app:factorization}

In Section~\ref{sec:4pt}, we explained how to bootstrap four-point functions in Grassmannian space. A key input was the factorization limit \eqref{equ:Res-S=0}:
\be \label{equ:Res-S=0X}
\mathop{\mathrm{Res}}_{S=0} A_4=\sum_hA_{3,L}^{(-h)} \hs A_{3,R}^{(+h)}\,,
\ee
where we have specialized to the $s$-channel.
In this appendix, we will show that this factorization property follows from unitarity and derive it from the cutting rule in momentum space.

\subsection{Cutting Rule} 

Our starting point is the momentum-space cutting rule for the exchange of a massless particle in the $s$-channel:
	\be\label{equ:cutting-app-alt}
	\mathop{\mathrm{Disc}}_{k_s^2}[\psi_4^{\hs \prime}] = - \frac{1}{2k_s}\sum_h \mathop{\mathrm{Disc}}_{k_s^2}[\psi_{3,L}^{\hs \prime \hs(-h)}]\,\mathop{\mathrm{Disc}}_{k_s^2}[\psi_{3,R}^{\hs \prime \hs (+h)}]\,,
	\ee
	where the discontinuity operation was defined in \eqref{equ:Disc-def}. Restoring the momentum-conserving delta function, this becomes
	\be
	\label{equ:Res-S-delta-function}
	\mathop{\mathrm{Disc}}_{k_s^2}[\psi_4] 
	=-\sum_h\int \frac{\ud^3 k_s}{2 k_s} \mathop{\mathrm{Disc}}_{k_s^2}[\psi_{3,L}^{\prime \hs (-h)}]\,\delta(\vec{k}_1+\vec{k}_2+\vec{k }_s) \hs \mathop{\mathrm{Disc}}_{k_s^2}[\psi_{3,R}^{\prime \hs(+h)}]\,\delta(\vec{k}_3+\vec{k}_4-\vec{k}_s) \,,
	\ee
	with $\psi_4 \equiv \delta(\vec{k}_1 + \vec{k}_2 + \vec{k}_3 + \vec{k}_4) \,\psi_4^{\hs \prime}$. To derive (\ref{equ:Res-S=0X}), we will now translate this cutting rule to Grassmannian space.

\vskip 4pt
We first write the discontinuities of the three-point functions in terms of their Grassmannian representations
\begin{align} 
\mathop{\mathrm{Disc}}_{k_s^2}[\psi_{3,L}^{(-h)}] &= \int \dif C_L \, \delta(C_L \cdot \Lambda_L) \, A_{3,L}^{(-h)}(C_L) \,, \\
\mathop{\mathrm{Disc}}_{k_s^2}[\psi_{3,R}^{(+h)}] &= \int \dif C_R \, \delta(C_R \cdot \Lambda_R) \, A_{3,R}^{(+h)}(C_R) \,, 
\end{align}
where $C_{L,R} \in \mathrm{OGr}(3,6)$. The matrices $\Lambda_L$ and $\Lambda_R$ are made from the spinors $\{\lambda^\alpha_{1,2,s}, \bar \lambda_{1,2,s}^\alpha\}$ and $\{\lambda^\alpha_{3,4,s}, \bar \lambda_{3,4}^\alpha, - \bar \lambda_s^\alpha\}$, respectively.  
 We can package the above data compactly as
\be
C_{L \times R} \equiv \left(\begin{matrix}
C_{L} &0\\
0&C_{R}
\end{matrix}\right), \quad \Lambda_{L \times R} \equiv \left(\begin{matrix}
\Lambda_L \\
\Lambda_R
\end{matrix}\right) .
\ee
Equation~\eqref{equ:Res-S-delta-function} can then be written as
\be
\label{eq:OGr612}
\mathop{\mathrm{Disc}}_{k_s^2}[\psi_4]  = -\int \dif C_{L \times R} \, \sum_h A_{3,L}^{(-h)} A_{3,R}^{(+h)} \int \frac{\ud^2\lambda_s \hs \ud^2\bar{\lambda}_s}{{\rm GL}(1)} \, \delta(C_{L \times R} \cdot \Lambda_{L \times R}) \,,
\ee
where the $k_s$-integral has been written in terms of the spinors $\lambda_s$ and $\bar \lambda_s$.
The final step is to perform the integral over the exchanged momentum by evaluating the delta function and to write the result in terms of the smaller Grassmannian matrix $C \in \mathrm{OGr}(4,8)$.
 
 \vskip 4pt
In Section~\ref{ssec:proof}, we will show that \eqref{eq:OGr612} can be written as
 \be
\label{eq:GrassDiscApp}
\mathop{\mathrm{Disc}}_{k_s^2}[\psi_4]  =  \mathop{\mathrm{Disc}}_{k_s^2} \left[ \int \dif C \: \delta(C \cdot \Lambda) \, A_4 \right] = -  \int \dif C \: \delta(C \cdot \Lambda) \, \delta(S)\, \sum_h A_{3,L}^{(-h)} A_{3,R}^{(+h)} \,.
\ee
where the integral is over $\mathrm{OGr}(4,8)$. Note that the right-hand side is localized at $S\equiv (\bar 1 \bar 2 12)=0$.
To interpret the left-hand side in Grassmannian space, we write it in a fixed gauge:
\beq
\label{equ:XYZ}
 \mathop{\mathrm{Disc}}_{k_s^2} \left[ \int \dif C \: \delta(C \cdot \Lambda) \, A_4 \right]
= \delta(\vec{k}_1 + \vec{k}_2 + \vec{k}_3+ \vec{k}_4) \mathop{\mathrm{Disc}}_{k_s^2} \left[\int \frac{\dif \tau}{2\pi i} \: \frac{R_4(c_{ij}(\tau))}{S(\tau)}\right],
\eeq
where $R_4 \equiv \mathop{\mathrm{Res}}_{S = 0} A_4$ is regular at $\tau = \tau_s, \bar \tau_s$ and we have assumed that the Grassmannian correlator $A_4$ has a simple pole at $S=0$. 
We note that  the parameters $c_{ij}(\tau)$ in~\eqref{equ:cijtau} are not affected by flipping the sign of the energy, $k_s \mapsto - k_s$. The Mandelstam $S(\tau)$, on the other hand, changes under this operation because it exchanges $\tau_s\leftrightarrow \bar\tau_s$. The net effect of the discontinuity is therefore to flip the $i \epsilon$ prescription of $S$. 
Hence, we get
\begin{align}
 \mathop{\mathrm{Disc}}_{k_s^2} \left[\int \frac{\dif \tau}{2\pi i} \: \frac{R_4(c_{ij}(\tau))}{S(\tau)}\right] &= \int \frac{\dif \tau}{2\pi i} \: R_4(c_{ij}(\tau)) \left( \frac{1}{S(\tau)+i\epsilon} - \frac{1}{S(\tau)-i\epsilon} \right)  \nonumber \\
 &= - \int \dif\tau \: R_4(c_{ij}(\tau)) \hs \delta(S(\tau))\, .
\end{align}
Inserting this into (\ref{equ:XYZ}), we find
\be
	\mathop{\mathrm{Disc}}_{k_s^2} \left[ \int \dif C \: \delta(C \cdot \Lambda) \, A_4 \right] = -\int \dif C \: \delta(C \cdot \Lambda) \, \delta(S)\, R_4 \,.
\ee
Comparing this to \eqref{eq:GrassDiscApp}, we then get  \eqref{equ:Res-S=0X},  as required.

\subsection{Completed Proof}
\label{ssec:proof}

To complete the above derivation, we now derive the expression \eqref{eq:GrassDiscApp} from \eqref{eq:OGr612}:
\be
\label{equ:start}
\mathop{\mathrm{Disc}}_{k_s^2}[\psi_4]  = -\int \dif C_{L \times R} \, \sum_h A_{3,L}^{(-h)} A_{3,R}^{(+h)} \int \frac{\ud^2\lambda_s \hs \ud^2\bar{\lambda}_s}{{\rm GL}(1)} \, \delta(C_{L \times R} \cdot \Lambda_{L \times R}) \,,
\ee
Consider the case where $C_L$ is in the right branch, while $C_R$ is in the left branch:
\be 
\label{equ:leftrightgauge}
C_L =\left(\begin{matrix}
	1&0&0&0&-c_{12}&-c_{1s}\\
	0&1&0&c_{12}&0&-c_{2s}\\
	0&0&1&c_{1s}&c_{2s}&0
\end{matrix}\right) , \quad
C_R =\left(\begin{matrix}
	0&-c_{34}&-c_{3\bar s}&1&0&0\\
	c_{34}&0&-c_{4 \bar s}&0&1&0\\
	c_{3 \bar s}&c_{4 \bar s}&0&0&0&1
\end{matrix}\right).
\ee 
Other cases can be treated similarly. Performing the momentum integral in \eqref{eq:OGr612} is trivial: one just integrates the delta functions that arise from the third and sixth row of $C_{L \times R}$. These impose the following equalities:
\be
\begin{aligned}
	\lambda_s & = c_{s1} \bar \lambda_1 + c_{s2} \bar \lambda_2 \,,\\[4pt]
	\bar \lambda_s & = c_{3 \bar s} \lambda_3 + c_{4 \bar s} \lambda_4 \,.
\end{aligned}
\ee
Substituting this into the remaining delta functions, we obtain
\be
	\int \ud^2\lambda_s \hs \ud^2\bar{\lambda}_s \, \delta(C_{L \times R} \cdot \Lambda_{L \times R}) = \delta(C \cdot \Lambda) \,,
	\label{equ:B15}
\ee
where $\Lambda$ only contains the spinors $\{\lambda_{1,2,3,4}^\alpha, \bar \lambda_{1,2,3,4}^\alpha\}$ of the four external particles and $C$ is the following $\mathrm{OGr}(4,8)$ matrix:
\be
\label{equ:Cmatrix}
C =\left(\begin{matrix}
	1&0&-c_{1s}c_{3 \bar s}&-c_{1s}c_{4 \bar s}&0&-c_{12}&0&0\\
	0&1&-c_{2s}c_{3 \bar s}&-c_{2s}c_{4 \bar s}&c_{12}&0&0&0\\
	0&0&0&-c_{34}&c_{1s}c_{3\bar s}&c_{2s}c_{3 \bar s}&1&0\\
	0&0&c_{34}&0&c_{1s}c_{4 \bar s}&c_{2s}c_{4 \bar s}&0&1
\end{matrix}\right) .
\ee
Notice that all matrices of this form are localized on the $s$-channel pole, since
\be
S=(\bar 1\bar 2 12)=c_{1s}c_{3 \bar s}c_{2s}c_{4\bar s}-c_{1s}c_{4\bar s}c_{2s}c_{3\bar s}=0\,,
\ee
which is a consequence of the ``factorization'' property
$c_{lr}=c_{l s}c_{r \bar s}$,
with $l=1,2$ and $r=3,4$. We see that the larger matrix $C_{L \times R} \in \mathrm{OGr}(3,6) \times \mathrm{OGr}(3,6)$ has effectively been projected to a smaller matrix $C \in \mathrm{OGr}(4,8)$, whose minors are related to the minors of $C_L$ and $C_R$ according to \eqref{equ:minors-factorized}, as is easy to verify for \eqref{equ:Cmatrix}. The first line of \eqref{equ:minors-factorized} then accounts for the sum over helicities.

\vskip 4pt
Using \eqref{equ:B15}, equation (\ref{equ:start}) becomes
\be
\label{equ:midpoint}
\mathop{\mathrm{Disc}}_{k_s^2}[\psi_4]  = -\int \frac{\dif C_{L \times R}}{\mathrm{GL}(1)} \, \delta(C \cdot \Lambda) \, \sum_h A_{3,L}^{(-h)} A_{3,R}^{(+h)} \,.
\ee
The final step is to rewrite the integration measure as a function of $C$. In the gauge~\eqref{equ:leftrightgauge}, we have
\be
	\int \frac{\dif C_{L \times R}}{\mathrm{GL}(1)} = \int \frac{\dif c_{12} \, \dif c_{1s} \, \dif c_{2s} \, \dif c_{34} \dif c_{3 \bar s} \, \dif c_{4 \bar s}}{\mathrm{GL(1)}} \,,
\ee
where the $\mathrm{GL(1)}$ factor can be absorbed by setting, for example, $c_{1s} = 1$. 
We then insert
\be
	1 = \int \dif^4 c_{l r} \: \delta(c_{lr} - c_{l s} c_{r \bar s})\,,
\ee 
and use the newly added delta functions to perform the integrals with respect to $c_{ls}$ and $c_{r \bar s}$. The computation is straightforward and yields
\be
	\int \frac{\dif C_{L \times R}}{\mathrm{GL}(1)} = \int \dif C \: \delta(S) \,.
\ee
Substituting this into~\eqref{equ:midpoint}, we finally get
\be
\mathop{\mathrm{Disc}}_{k_s^2}[\psi_4]  = -  \int \dif C \: \delta(C \cdot \Lambda) \, \delta(S)\, \sum_h A_{3,L}^{(-h)} A_{3,R}^{(+h)}  \,,
\ee
as required.

\newpage
\section{Recovering Flat-Space Physics}
\label{app:flat}
\begin{figure}[h!]
	\centering
	\includegraphics[scale=0.5]{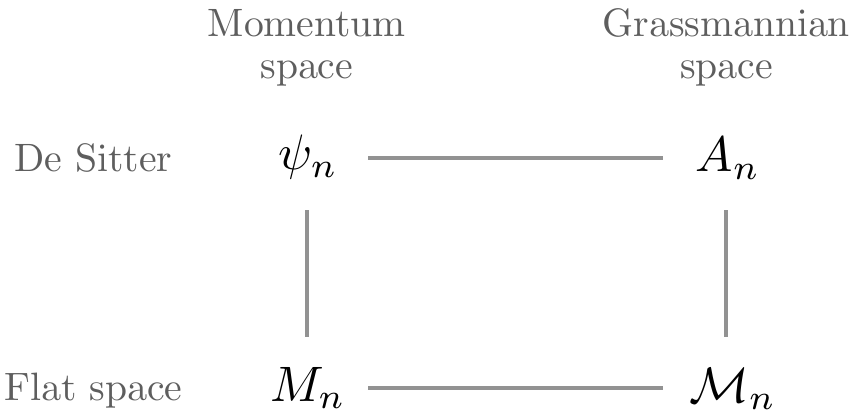} 
	\caption{In this appendix, we describe the flat-space limit in Grassmannian space, in which the correlator $A_n$ reduces to the amplitude ${\cal M}_n$. We further show that, in this limit, the Grassmannian integral  produces the flat-space amplitude $M_n$. This is complementary to the treatment in the main text, where the Grassmannian integral gives the wavefunction coefficient $\psi_n$ and the flat-space limit, which leads to the amplitude $M_n$, is taken in momentum space.}
	\label{fig:FlatSpace}
\end{figure}

In this appendix, we discuss the flat-space limit of our Grassmannian correlators.
We will first identify the limit in which the cosmological Grassmannian reduces to the flat-space Grassmannian,\footnote{We thank Yu-tin Huang for insightful discussions that inspired the approach presented in this appendix.} and then show that in the same limit the momentum-space correlator has a total energy singularity whose coefficient is the corresponding flat-space scattering amplitude (see Figure~\ref{fig:FlatSpace}). 

\subsection{Flat-Space Grassmannian}

Amplitudes of conformal theories can be written as~\cite{Arkani-Hamed:2009hub, Arkani-Hamed:2009ljj} 
\beq
\label{eq:amplitudegrass}
M_n(\mathcal{L}, \tilde{\mathcal{L}}) = \int \frac{\ud^{k \times n} {\cal C}}{\mathrm{GL}(k)} \, \delta({\cal C} \cdot \tilde{\mathcal{L}}) \, \delta({\cal C}^\perp \cdot \mathcal{L}) \, {\cal M}_n({\cal C}) \,,
\eeq
where the integral is over the Grassmannian $\mathrm{Gr}(k,n)$, for some $2 \leq k \leq n-2$.
The integrand $\mathcal{M}_n(\mathcal{C})$ depends on the minors of $\mathcal{C}$ and transforms as $\mathcal {M}_n\to \rho^{4-n}\hs \mathcal {M}_n$ under rescalings of each minor by a factor of $\rho$. 
The value of $k$ is a function of the specific amplitude under consideration. For example, in pure Yang--Mills theory, $k$ is the number of external gluons with negative helicity. The $n \times 2$ matrices $\mathcal{L}$ and $\tilde{\mathcal{L}}$ contain the flat-space spinors $\lambda_i^\alpha$ and $\tilde \lambda_i^{\dot \alpha}$, respectively, and four-momentum conservation implies $\mathcal{L}^T \cdot \tilde{\mathcal{L}} = 0$. The matrix ${\cal C}^\perp$ is the $(n-k)$-dimensional complement of ${\cal C}$, i.e.~${\cal C}^\perp \cdot {\cal C}^T = 0$. Even when the theory under study is not conformal, but simply Poincaré-invariant, a representation like \eqref{eq:amplitudegrass} still holds, but the integrand $\mathcal{M}_n$ then has an additional dependence on spinor brackets.

\vskip 4pt
We will now identify a limit in which the orthogonal Grassmannian $\mathrm{OGr}(n,2n)$ reduces to the flat-space Grassmannian  $\mathrm{Gr}(k,n)$.  First, we write the orthogonal Grassmannian as $C = \big(C^{(1)}_{n \times n},  C^{(2)}_{n \times n} \big)$. 
The constraint of momentum conservation, $C \cdot \Lambda = 0$, then becomes 
\beq
C^{(1)} \cdot L + C^{(2)} \cdot \bar L = 0 \,,
\label{equ:mix}
\eeq
where $L$ and $\bar L$ have the same meaning as in \eqref{eq:amplitudegrass}, but only satisfy three-momentum conservation, i.e.~$L^T \cdot \bar{L} \ne 0$. We are looking for a limit $C^{(i)} \to C^{(i)}_{\rm flat}$ in the Grassmannian 
corresponding to the flat-space kinematics $L \to \mathcal{L}, \bar L \to \tilde{\mathcal{L}}$, where the theory acquires an emergent Poincaré symmetry. This demands that dotted and undotted spinors do not mix, so that the two terms in (\ref{equ:mix}) should vanish independently:
\begin{equation}
	\label{eq:flatlocalization}
	C^{(1)}_{\rm flat} \cdot \mathcal{L} = 0 \,, \quad C^{(2)}_{\rm flat} \cdot \tilde{\mathcal{L}} = 0 \,.
\end{equation}
Comparing this to \eqref{eq:amplitudegrass},  it is natural to impose the isomorphism $C^{(1)}_{\rm flat} \simeq {\cal C}^\perp$ and $C^{(2)}_{\rm flat} \simeq {\cal C}$. We must therefore further demand that the rank of $C^{(2)}_{\rm flat}$ equals $k$, which is determined by the target amplitude. Equation~\eqref{eq:flatlocalization} is consistent with an emergent four-momentum conservation, as one can write $\mathcal{L} = \mathcal{C}^T \cdot P$, for some $k \times 2$ matrix $P$,\footnote{By definition, the kernel  of $\mathcal{C}^\perp$ must be expressible in terms of the rows of $\mathcal{C}$.} and substitute this into $\mathcal{L}^T \cdot \tilde{\mathcal{L}} = P^T \cdot (\mathcal{C} \cdot \tilde{\mathcal{L}}) = 0$. This representation of $\mathcal{L}$ also implies $k \geq 2$, since for $k = 1$ its columns would be proportional to each other.

\vskip 4pt
Since the rank of the matrices $C^{(i)}_{\rm flat}$ is at most $n-2$ (with $n=3$ being a special case), which further enforces $k \leq n-2$, it automatically follows that their determinants vanish 
\beq
\begin{aligned}
	\label{eq:necesflat}
	\det C^{(1)}_{\rm flat} & \equiv (\bar 1\bar2 \cdots \bar n) = 0 \,,\\ 
	\det C^{(2)}_{\rm flat} & \equiv (12\cdots n) = 0 \,.
\end{aligned}
\eeq
Finally, we observe that a $\mathrm{GL}(n)$ transformation in $\mathrm{OGr}(n,2n)$ acts as a $\mathrm{GL}(k)$ transformation on ${\cal C}$ and as a $\mathrm{GL}(n-k)$ transformation on ${\cal C}^\perp$, hence matching the ``gauge group'' in \eqref{eq:amplitudegrass}.

\vskip 4pt
In short, the above analysis has taught us that the flat-space limit can be extracted from $\mathrm{OGr}(n,2n)$ by imposing $C^{(2)}$ to be an element of $\mathrm{Gr}(k,n)$ and $C^{(1)}$ to be an element of its complement $\mathrm{Gr}(n-k,n)$. Moreover, since $\mathrm{Gr}(k,n)$ is a connected geometry, it must localize entirely in one branch of $\mathrm{OGr}(n,2n)$, which is expected to depend on $k$.

\subsection{Total Energy Singularities}

In the following, we will further illustrate these consideration for the special case $n=4$.
We will show that the flat-space limit \eqref{eq:necesflat} 
corresponds to the limit $S+T+U \to 0$ in Grassmannian space  and the limit $E \to 0$ in momentum space. It therefore relates to the well-known total energy singularities of correlators in momentum space. 

\vskip 4pt 
We first note that
\beq
(S+T+U)^2 = 4\hs (\bar 1 \bar 2 \bar 3 \bar 4)(1234) \,,
\eeq
so that \eqref{eq:necesflat} implies $S+T+U=0$, as required.
To prove the second claim, it is convenient to pick the following gauge:\footnote{We do not need to consider the left branch of $\mathrm{OGr}(4,8)$, as \eqref{eq:flatlocalization} demands the rank of $C^{(1)}$ to be $2$ and $C^{(1)}_{\rm flat}$ to be an element of $\mathrm{Gr}(2,4)$ (recall that $2 \leq k \leq n-2$), which is only possible in the right branch.}
\beq
\label{eq:appendixgauge}
C = \begin{pmatrix}
	0 & -\tilde c_{12} & 0 & 0 & 1 & 0 & -\tilde c_{13} & -\tilde c_{14} \\
	\tilde c_{12} & 0 & 0 & 0 & 0 & 1 & -\tilde c_{23} & -\tilde c_{24} \\
	\tilde c_{13} & \tilde c_{23} & 1 & 0 & 0 & 0 & 0 & -\tilde c_{34} \\
	\tilde c_{14} & \tilde c_{24} & 0 & 1 & 0 & 0 & \tilde c_{34} & 0
\end{pmatrix} ,
\eeq
so that
$(\bar 1\bar 2 \bar3 \bar4) = \tilde c_{12}^2$ and $(1234) = \tilde c_{34}^2$.  We therefore see explicitly that the condition \eqref{eq:necesflat} implies $\tilde c_{12} = \tilde c_{34} = 0$
and hence
\beq
C \to C_{\rm flat} = \begin{pmatrix}
	0_{2 \times 4} & {\cal C}_{2 \times 4} \\
	{\cal C}^\perp_{2\times 4} & 0_{2 \times 4}
\end{pmatrix} ,
\eeq
which is the structure demanded by \eqref{eq:flatlocalization}. In this limit, the conditions \eqref{eq:flatlocalization} imply $\mathcal{C} = \mathcal{L}$ and $\mathcal{C}^\perp = \tilde{\mathcal{L}}$. An obvious consequence is that only minors of the type $(i j \bar k \bar l)$ are non-zero, which evaluate in momentum space to $\langle i j \rangle [k l]$.

\vskip 4pt
We would now like to see what the condition $\tilde c_{12} = \tilde c_{34} = 0$ translates to in momentum space. Solving the constraint $C \cdot \Lambda = 0$ in the gauge~\eqref{eq:appendixgauge} yields the same result as \eqref{equ:cijtau}, up to an interchange of $\lambda_{1,2} \leftrightarrow \bar \lambda_{1,2}$:
\beq
\begin{aligned}
	\tilde c_{12}(\tilde \tau) & = \frac{\langle \bar 1 \bar 2 \rangle}{E- 2k_1 - 2k_2} + \tilde \tau \langle \bar 3 \bar 4 \rangle \,, \\[4pt]
	\tilde c_{34}(\tilde \tau) & = \frac{\langle 3 4 \rangle}{E- 2k_1 - 2k_2} + \tilde \tau \langle 1 2 \rangle \, .
\end{aligned}
\eeq
Setting both of these to $0$, and using some spinor identities, leads to
\be
E = 0\quad {\rm and} \quad 
\tilde \tau =  -\frac{\langle 3 4 \rangle}{\langle 1 2 \rangle} \frac{1}{E - 2k_1-2k_2} \equiv \tilde \tau_0\,.
\ee
As expected, we recover the condition $E = 0$. To make sense of the result for $\tilde \tau$, we perform a $\mathrm{GL}(n)$ transformation that relates the parameters $\tilde c_{ij}(\tilde \tau)$ in the gauge \eqref{eq:appendixgauge} to the parameters $c_{ij}(\tau)$ in the gauge \eqref{equ:C-slice}. We find that
\beq
\frac{1}{\tilde c_{12}(\tilde \tau)}  = -  c_{12}(\tau) = - \frac{\langle 1 2 \rangle}{E} - \tau \langle \bar 3 \bar 4 \rangle \,.
\eeq
Solving $\tau$ in terms of $\tilde \tau_0$ then leads to $\tau(\tilde \tau_0) = 0$. 
The Grassmannian flat-space limit therefore predicts that the leading contribution as $E \to 0$ arises entirely from the $\tau = 0$ pole. Indeed, it is straightforward to verify that all coefficients $c_{ij}(\tau)$ in \eqref{equ:cijtau} remain finite as $E \to 0$ when evaluated on any of the other poles contributing to the correlator.

\subsection{Amplitudes from Singularities}

It is well known that the coefficient of the total energy singularity of the correlator $\psi_n$ is the corresponding flat-space amplitude $M_n$. In the previous section, we showed how this limit can be taken directly in Grassmannian space. To complete the story, we will now demonstrate how to extract the Grassmannian amplitude  ${\cal M}_n$ from the flat-space limit of the Grassmannian correlator~$A_n$.
We will first present a general proof of this relationship and then give a few explicit examples.

\subsubsection*{Proof}

We start from the Grassmannian integral of the correlator:
\beq
\psi_n(\Lambda) = \int \frac{{\rm d}^{n\times 2n}C}{\text{GL}(n)} \, \delta(C\cdot Q\cdot C^T)\, \delta(C\cdot \Lambda) \,A_n(C)\,.
\label{equ:INTEGRAL}
\eeq
We want to show explicitly how, in the flat-space limit, this transforms into the integral (\ref{eq:amplitudegrass}), 
and then extract the flat-space Grassmannian amplitude ${\cal M}_n$ from the leading singularity of the de Sitter Grassmannian correlator $A_n$.

\vskip 4pt
We will choose the following parameterization of the orthogonal Grassmannian OGr$(n,2n)$:
\be \label{equ:parametr-n-k}
C=\begin{pmatrix}- c_{ij}& 0_{k\times (n-k)}& 1_{k\times k}&-c_{iJ}\\-c_{Ij}& 1_{(n-k)\times (n-k)} & 0_{(n-k)\times k}&-c_{IJ}\end{pmatrix} ,
\ee 
where we have set the matrix with columns $1,2,\cdots\hskip -1pt,k,\overline{k+1},\overline{k+2},\cdots\hskip -1pt,\bar n$ to be the identity. This parameterizes the right or left branch of OGr$(n,2n)$ if $k$ is even or odd, respectively.  We have split the
non-trivial components of the Grassmannian into two anti-symmetric matrices $c_{ij}$ and $c_{IJ}$, with $i,j=1,2,\cdots\hskip -1pt,k$ and $I,J=k+1,\cdots\hskip -1pt,n$, as well as a rectangular matrix~$c_{iJ}$, with $c_{Ji}=-c_{iJ}$.
The delta function $\delta(C\cdot \Lambda)$ in the integral~\eqref{equ:INTEGRAL} can then be written as
\begin{align} \label{equ:delta-halfFT}
\delta(C\cdot \Lambda)&=\ \delta\big(\bar\lambda_i-c_{iJ}\bar\lambda_J -c_{ij}\lambda_j\big)\,\delta\big(\lambda_I-c_{Ij}\lambda_j -c_{IJ}\bar\lambda_J\big)\nonumber\\[6pt]
&=\ \int \ud^2\mu_i \hs \ud^2\bar\mu_I \,\exp\Big(i\hs \bar\lambda_i\cdot \mu_i-i \lambda_I\cdot \bar\mu_I\Big)\hs \exp\Big(\hskip -2pt-i\hs  c_{iJ}\hs Z_i\cdot W_J\Big)\nonumber\\
&\quad\quad\times\exp\left(
\frac{i}{2}\hs  c_{ij}\hs Z_i\cdot Z_j -\frac{i}{2} \hs c_{IJ}\hs W_I\cdot W_J \right) ,
\end{align} 
where repeated indices are implicitly summed over. Since the spacetime is specified by the infinity twistor used in the contractions $Z_i\cdot Z_j $ and $W_I\cdot W_J$, taking the flat-space limit amounts to replacing the (A)dS infinity twistor $I_{\rm (A)dS} = \Omega$ with its flat-space counterpart $I_{\rm flat}$, so that
\be 
\begin{aligned}
 Z_i\cdot Z_j &\mapsto  L\hs \langle ij\rangle \,,\\
   W_I\cdot W_J &\mapsto L\hs [IJ]\,,
   \end{aligned}
\ee 
where the flat-space spinor brackets are $\langle ij\rangle \equiv Z_i\cdot I_{\rm{flat}}\cdot Z_j$ and $[IJ] \equiv W_I\cdot I_{\rm{flat}}\cdot W_J$. 
We restored the (A)dS radius $L$, which tends to $\infty$ in the flat-space limit, to keep $Z_i\cdot Z_j$ and $W_I\cdot W_J$ dimensionless.  Furthermore, we took $\bar\lambda^\alpha \to \tilde\lambda^{\dot\alpha}$, with a dotted spinor index, inside the square spinor brackets $[IJ]$.
The exponential in the last line of \eqref{equ:delta-halfFT} can then be factored out the integrals, and the delta function reads 
\be
\delta(C\cdot\Lambda)\mapsto \delta\big(\tilde\lambda_i-c_{iJ}\tilde\lambda_J \big)\hs \delta\big(\lambda_I+c_{jI}\lambda_j \big)\hs\exp\left(\frac{i}{2}\hs  c_{ij}\hs L\hs \langle ij\rangle -\frac{i}{2} \hs c_{IJ}\hs L\hs [IJ]\right).
\ee
Note that dotted and undotted indices do not mix, as expected in flat space.
Plugging this back into~\eqref{equ:INTEGRAL}, we get 
\be \label{equ:delta-flatspace-param}
M_n(\mathcal{L}, \tilde{\mathcal{L}}) =\int \ud c_{iJ} \, \delta\big(\tilde\lambda_i-c_{iJ}\tilde\lambda_J \big)\hs \delta\big(\lambda_I+c_{jI}\lambda_j \big)\,  \mathcal{M}_n(c_{iJ},\langle ij\rangle,[IJ] )\,,
\ee 
where the integrand is 
\be \label{equ:Mn-integral}
\mathcal{M}_n(c_{iJ},\langle ij\rangle,[IJ] )
=\lim_{L\to \infty}\int \ud c_{ij}\hs \ud c_{IJ}\hs A_n(c_{iJ},c_{ij}, c_{IJ}) \hs \exp\left(\frac{i}{2} c_{ij}\hs L\hs \langle ij\rangle -\frac{i}{2} c_{IJ}\hs L\hs [IJ]\right) .
\ee 
The integral \eqref{equ:delta-flatspace-param} is equal to the flat-space Grassmannian integral~\eqref{eq:amplitudegrass} after parameterizing Gr$(k,n)$ as $\mathcal{C}=\big( 1_{k\times k} \,\hs ,\hs -c_{iJ}\big)$. 

\vskip 4pt
To relate $\mathcal{M}_n$ to the leading singularity of $A_n$, we assume that
\beq
\lim_{c_{ij},c_{IJ}\to0 } A_n \sim \prod_{i<j,I<J}    c_{ij}^{-1-r_{ij}} c_{IJ}^{-1-r_{IJ}} \,,
\eeq
where $r_{ij}$ and $r_{IJ}$ are non-negative integers. 
In the flat-space limit, we isolate the contribution of the integral arising from the pole at $c_{ij} = c_{IJ} = 0$. This implies that, in the limit $c_{ij},c_{IJ}\to0$, the leading singularity of $A_n$ determines $\mathcal{M}_n$
 in \eqref{equ:Mn-integral} according to
\be \label{equ:Mnmathcal}
\mathcal{M}_n(c_{iJ},\langle ij\rangle,[IJ] ) =\lim_{c_{ij},c_{IJ}\to0 }\prod_{i<j,I<J} \langle ij\rangle^{r_{ij}} \hs [IJ]^{r_{IJ}}\hs c_{ij}^{1+r_{ij}}\hs c_{IJ}^{1+r_{IJ}}A_n(c_{iJ},c_{ij}, c_{IJ})\,,
\ee 
where we dropped a constant factor that depends on the scale $L$ and a sign that is determined by the parameters $r_{ij}, r_{IJ}$. In the following, we will confirm this result in a few explicit examples.

\subsubsection*{Examples}

For concreteness, we will focus on the cases $n = 3$ and $n=4$. We will first consider pure Yang--Mills theory, and then the case of non-conformal bulk theories where the flat-space Grassmannian amplitude $\mathcal{M}_n$ depends explicitly on spinor brackets $\langle ij\rangle$, $[IJ]$.

\begin{itemize}
	\item For $n = 3$, we parameterize the right branch in the following gauge:
	\beq
	\label{eq:appgaugeflat3}
	C = \begin{pmatrix}
		0 & - c_{12} & 0 & 1 & 0 & - c_{13}\\
		c_{12} & 0 & 0 & 0 & 1 & - c_{23}\\
		 c_{13} & c_{23} & 1 & 0 & 0 & 0
	\end{pmatrix} ,
	\eeq
	so that the $--+$ correlator in Yang--Mills theory takes the form
	\beq
	A(1^- 2^- 3^+) = \frac{1}{c_{12} c_{23} c_{31}} \,.
	\eeq
	According to \eqref{equ:Mnmathcal}, the flat-space Grassmannian amplitude  then is 
	\beq\label{equ:mathcalM3}
	\mathcal{M}(1^- 2^- 3^+) = -\lim_{ c_{12}\to0}  c_{12}\hs A(1^- 2^- 3^+) = \frac{1}{c_{23} c_{13}} \,.
	\eeq
	Indeed, given the Gr$(2,3)$ matrix
	\beq
	{\cal C} = \begin{pmatrix}
		1 & 0 & - c_{13} \\
		0 & 1 & - c_{23}
	\end{pmatrix},
	\eeq
	we can uplift~\eqref{equ:mathcalM3} to an expression in terms of the minors $(i j)$ of~$\mathcal{C}$ as
	\beq
		\mathcal{M}(1^- 2^- 3^+) 
		= \frac{(12)^4}{(12)(23)(31)}\,.
	\eeq
	Evaluating the integral \eqref{eq:amplitudegrass} simply amounts to the substitution $(i j) \mapsto \langle i j \rangle$, as the delta functions impose $\mathcal{C} = \mathcal{L}$.
	This confirms that we correctly recover the Yang--Mills amplitude.

	\item For $n = 4$, we express the $--++$ reduced Yang--Mills correlator in the gauge \eqref{eq:appendixgauge} as
		\begin{align}
		A(1^- 2^- 3^+ 4^+) & =  2 \left(\frac{1}{S+T+U} + \frac{1}{S+T-U}\right) \frac{(1 2 \bar 3 \bar 4)^2}{S T} \nonumber \\[4 pt]
		& = \left( \frac{1}{c_{12} c_{34}} - \frac{1}{ c_{14}  c_{23}} \right) \frac{1}{(c_{13} c_{24} -  c_{14} c_{23}) (c_{12} c_{34} - c_{13} c_{24})} \,.
	\end{align}
According to \eqref{equ:Mnmathcal}, the flat-space Grassmannian amplitude then is 
	\beq
	\mathcal{M}(1^- 2^- 3^+ 4^+) = -\lim_{c_{12}, c_{34}\to0} c_{12}\hs c_{34}\hs A(1^- 2^- 3^+ 4^+) = \frac{1}{c_{13} c_{24} (c_{13} c_{24} - c_{14} c_{23})} \,.
	\label{equ:MMn}
	\eeq
	Given the Gr$(2,4)$ matrix
	\beq
	{\cal C} = \begin{pmatrix}
		1 & 0 & - c_{13} & - c_{14} \\
		0 & 1 & - c_{23} & - c_{24}
	\end{pmatrix},
	\eeq
	we can uplift \eqref{equ:MMn} to an expression in terms of the minors $(ij)$ of $\mathcal C$ as
	\beq
		\mathcal{M}(1^- 2^- 3^+ 4^+)  = \frac{(12)^4}{(12)(23)(34)(41)} \,.
	\eeq
	 Computing the integral \eqref{eq:amplitudegrass} by the substitution $(ij) \mapsto \langle i j\rangle$, we again recover the correct Yang--Mills amplitude. Performing the same computation with the full color-ordered correlator \eqref{equ:AYM-kk} gives the same result, up to an overall factor of $3/2$.
\end{itemize}

\vskip 4pt
We note that, in the case of  Yang--Mills theory, the poles of $A_n$ at $c_{ij},c_{IJ}\to0$ are always simple, and thus the factors of $\langle ij\rangle$, $[IJ]$ in~\eqref{equ:Mnmathcal} are trivial. This leads to a Grassmannian amplitude~$\mathcal M_n(\mathcal C)$ with no explicit dependence on spinor brackets. 
This is related to the Yang--Mills interactions being conformal in the bulk.
We will now show examples of non-conformal theories where the Grassmannian amplitude inherits a dependence on the spinor brackets~$\langle ij\rangle$ and $[IJ]$.

\begin{itemize}
	\item For $n = 3$, we study as an example the $--+$ graviton correlator. In the gauge \eqref{eq:appgaugeflat3}, it takes the form
	\beq
	A(1^{--} 2^{--} 3^{++}) = \frac{1}{(c_{12} c_{23} c_{13})^2} \,.
	\eeq
The singularity at $c_{12}=0$ is now a double pole, and thus equation~\eqref{equ:Mnmathcal} implies that the Grassmannian amplitude is given by 
	\begin{align}
	\mathcal{M}(1^{--} 2^{--} 3^{++}) & = \lim_{c_{12}\to0} \langle12\rangle \hs c_{12}^2\hs A(1^{--} 2^{--} 3^{++})= \frac{\langle 12 \rangle}{(c_{23} c_{13})^2}\nonumber \\[4pt]
	&  = \langle 1 2 \rangle \frac{(12)^7}{(12)^2 (23)^2 (31)^2} \,,
	\end{align}
	where we uplifted the result to an expression in terms of minors of $\mathcal C$. Computing the integral \eqref{eq:amplitudegrass} by the substitution $(i j) \mapsto \langle i j \rangle$ returns the correct graviton amplitude.
	
	\item For $n = 4$, we study as an example the graviton exchange \eqref{equ:Ag}. In the gauge \eqref{eq:appendixgauge}, it takes the form
	\beq
	A_{4,g} = \frac{1}{4(c_{12}  c_{34})^2} \left( \frac{(c_{14}  c_{23} +  c_{13}  c_{24})^2}{c_{13} c_{24} - c_{14}  c_{23}} - \frac{c_{13} c_{24} - c_{14} c_{23}}{3} \right).
	\eeq
Now, equation~\eqref{equ:Mnmathcal} implies that 
the Grassmannian amplitude is given by 
	\begin{align}
		\mathcal{M}_{4,g} & =\lim_{c_{12},c_{34}\to0}\langle 12\rangle\hs [34]\hs c_{12}^2\hs c_{34}^2\hs A_{4,g}  \nonumber \\[4pt]
		& = \frac{\langle 1 2 \rangle [3 4]}{4} \left( \frac{(c_{14} c_{23} + c_{13} c_{24})^2}{c_{13} c_{24} - c_{14} c_{23}} - \frac{c_{13} c_{24} - c_{14}  c_{23}}{3} \right) \nonumber \\[4pt]
		& = \frac{\langle 1 2 \rangle [34]}{4(12)^2} \left( \frac{((13)(42)-(14)(23))^2}{(12)(34)} - \frac{(12)(34)}{3} \right) ,
	\end{align}
	where we again uplifted the result to an expression in terms of minors of $\mathcal C$.
	Performing the substitution~$(ij) \mapsto \langle i j \rangle$, and using spinor identities, yields
	\beq
	M_{4,g} = \frac{1}{4} \left( \frac{(u-t)^2}{s} - \frac{s}{3} \right),
	\eeq
	which is the correct graviton exchange amplitude. 
\end{itemize}

\newpage
\section{Graviton Correlator}
\label{app:gravity}

The tree-level scattering amplitude of four gravitons is described by a very compact formula~\cite{Cheung:2017pzi}: 
\be
M(1^- 2^- 3^+ 4^+) = -\frac{\langle 1 2 \rangle^4 [3 4]^4}{s \hs t \hs u} \,.
\label{eq:posterchild2}
\ee
In contrast, the corresponding four-point graviton correlator (in momentum space) does not showcase the same level of simplicity~\cite{Bonifacio:2022vwa}. In this appendix, we demonstrate that the four-point function in Grassmannian space is dramatically simpler. 

\subsection{Grassmannian Space}

Like for the Yang--Mills correlator, we sum the contributions from multiple channels. Using the factorization property~\eqref{equ:Res-S=0} and the results of Section~\ref{sec:gaugeinv3pt}, the residue of the $s$-channel pole is
\begin{align}
	\mathop{\rm{Res}}_{S=0}A(1^- 2^- 3^+ 4^+) &= \Bigg(\frac{(1  2\bar I_s)^2}{(2 I_s\bar I_s)(I_s \bar 2\bar I_s)}\hs \frac{(I_s \bar 3\bar 4)^2}{(3 \bar 4 4)(4 \bar 3\bar 4)} \Bigg)^2 + \Bigg(\frac{(1 2 I_s)^2}{(2 \bar I_s I_s)(I_s \bar 2\bar I_s)}\hs \frac{(\bar I_s \bar 3\bar 4)^2}{(3 4\bar 4)(4 \bar 3\bar 4)} \Bigg)^2 \nonumber\\[4pt]
	&=\frac{(1 2\bar 3 \bar4)^4}{(2 \bar3 \bar 4 4)^2(\bar2 3\bar 44)^2}\bigg|_{S=0}=\frac{16\hs ( 1 2\bar 3 \bar4)^4}{(S+T+U)^2(T-U)^2}\bigg|_{S=0}\,.
\end{align}
To arrive at the expression in the second line, we have added vanishing mixed terms involving $(12\bar I_s)(I_s\bar3\bar4)$ and $(12 I_s)(\bar I_s\bar3\bar4)$. 
Similarly, for the $t$- and $u$-channels, we get 
\be
\begin{aligned}
	\mathop{\rm{Res}}_{T=0}A(1^- 2^- 3^+ 4^+) & = \frac{16\hs ( 1 2\bar 3 \bar4)^4}{(S+T+U)^2(S-U)^2}\bigg|_{T=0}\,, \\[4pt]
	\mathop{\rm{Res}}_{U=0}A(1^- 2^- 3^+ 4^+) & = \frac{16\hs ( 1 2\bar 3 \bar4)^4}{(S+T+U)^2(T-S)^2}\bigg|_{U=0}\,.
\end{aligned}
\ee
We observe that all residues are obtained from squaring the residues of the Yang--Mills result, which simply reflects the double-copy relation of the three-point functions.

\vskip 4pt
As before, we need to extend the correlator away from these factorization limits. We proceed as follows. In Yang--Mills theory, there are two structures that are consistent with the factorization on $S = 0$ and contain a singularity at $S+T+U=0$, namely
\be
\mathop{\rm{Res}}_{S=0} A_{4,{\rm YM}}^{(1)} = \frac{4\hs (12 \bar 3 \bar 4)^2}{(S+T+U)(S+T-U)} \,, \quad \mathop{\rm{Res}}_{S=0} A_{4,{\rm YM}}^{(2)} = -\frac{4\hs (12\bar3 \bar 4)^2}{(S+T+U)(S-T+U)} \,.
\ee
The color-ordered correlator \eqref{equ:AYM-kk} was obtained by adding these two contributions. 
For gravity, we consider an ansatz in which the squares of these two structures contribute with equal weight, as required by Bose symmetry, while omitting mixed terms arising from their product.
This gives\footnote{While we prepared V2 of this paper, this formula also appeared in~\cite{Bala:2026lvw}. Here, we provide an additional check of this ansatz by computing the associated momentum-space correlator (see Section~\ref{sec:GR-momentum}).}
\beq
\begin{aligned}
	A(1^- 2^- 3^+ 4^+)  = 8\,\frac{(12\bar 3\bar 4)^4}{(S+T+U)^2} \Bigg[ &\frac{1}{S} \left( \frac{1}{(S+T-U)^2} + \frac{1}{(S-T+U)^2}\right)  \\
	+\, &\frac{1}{T} \left(\frac{1}{(S+T-U)^2} + \frac{1}{(-S+T+U)^2}\right) \\
	 +\, &\frac{1}{U} \left(\frac{1}{(S-T+U)^2} + \frac{1}{(-S+T+U)^2} \right)\Bigg]\, .
\end{aligned}	 
	\label{eq:fourgravitons-0}
\eeq
To reveal a similar structure as in~\eqref{eq:posterchild2}, it
is nice to write the result as
\be
\hspace{-0.15cm}\boxed{A(1^- 2^- 3^+ 4^+) = -2 \frac{(1 2 \bar 3 \bar 4)^4}{S T U}  \Bigg[ \frac{3}{(S+T+U)^2} - \bigg(\frac{1}{(S+T-U)^2} + (T \leftrightarrow U) + (U  \leftrightarrow S) \bigg) \Bigg] }
\label{eq:fourgravitons}
\ee
As for the Yang--Mills correlator, the answer is a deformation of the flat-space amplitude. Indeed, the flat-space limit of \eqref{eq:fourgravitons} leads to
\be 
\lim_{\mathcal{E}\to0} A(1^- 2^- 3^+ 4^+)= \lim_{\mathcal{E}\to0} -\frac{1}{\mathcal{E}^2} \hs \frac{3}{2} \hs\frac{(1 2 \bar 3 \bar 4)^4}{S T U}   \mapsto \frac{3}{E^3}\bigg(\hspace{-3pt} - \frac{\langle 12\rangle^4[34]^4}{s\hs t \hs u}\bigg)\,,
\ee 
which contains the correct graviton amplitude~\eqref{eq:posterchild2}.

\subsection{Momentum Space}
\label{sec:GR-momentum}

We briefly comment on the relation to the answer in momentum space; cf.~equation (3.21) in~\cite{Bonifacio:2022vwa}.
As before, the momentum-space result is obtained by performing the integral \eqref{equ:tpsi-int4}.
This time, we can choose a contour that produces a triple discontinuity of the wavefunction coefficient:
\be
\int \frac{\ud \tau}{2\pi i} \: A(1^- 2^- 3^+ 4^+) = \mathop{\rm{Disc}}_{k_1^2,k_2^2,k_3^2}\hs\big[\psi(1^- 2^-3^+ 4^+)\big]\,.
\label{equ:D8}
\ee
Concretely, this result is obtained by adding up $2/3$ of the residue at $\tau=0$ plus the residues at $\tau=\bar\tau_s,\bar\tau_t,\tau_{12},\tau_{23},\tau_{13}$. Different choices of residues $\tau_{ij}$ yield all the other triple discontinuities. Although this contour choice deserves to be better understood, it is worth noting that the same contour also reproduces the triple discontinuity of the color-ordered Yang--Mills correlator in momentum space from the Grassmannian correlator \eqref{equ:AYM-kk}. The momentum-space result (\ref{equ:D8}) is therefore a nontrivial check of the ansatz (\ref{eq:fourgravitons}).

\vskip 4pt
Note that double discontinuities cannot be extracted from \eqref{eq:fourgravitons}. Quadruple discontinuities, on the other hand, can be obtained systematically by encircling $\bar\tau_{s,t,u}$ counter-clockwise and $\tau_{s,t,u}$ clockwise. However, these discontinuities miss important physical information, such as the $E \to 0$ limit. We leave a more systematic investigation of these issues to future work.

\newpage
\phantomsection
\addcontentsline{toc}{section}{References}
\bibliographystyle{utphys}
{\linespread{1.075}
	\bibliography{Magic-Refs}
}

\end{document}